\documentclass[prd,nofootinbib,floats,aps,twocolumn,tightenlines,superscriptaddress]{revtex4-2}
\usepackage{amsmath}
\usepackage{amssymb}
\usepackage{amsfonts}
\usepackage{mathrsfs}
\usepackage{graphicx}
\usepackage[caption=false]{subfig}
\usepackage{enumitem}
\usepackage[normalem]{ulem}
\usepackage[percent]{overpic}
\usepackage{bm}
\usepackage{rotating}
\usepackage[usenames, dvipsnames]{color}
\usepackage{hyperref}
\hypersetup{
    colorlinks=true,
    linkcolor=blue,
    filecolor=magenta,      
    urlcolor=cyan,
}
\usepackage{cancel}
\usepackage{verbatim}
\usepackage{mathtools}
\usepackage{graphicx}
\usepackage{tensor}
\usepackage{verbatim}
\usepackage{tikz}
\usepackage{booktabs}
\graphicspath{ {images/} }

\usepackage{braket}

\newcommand{\ii}{\mathrm{i}}

\renewcommand{\d}{\mathrm{d}}

\newcommand{\be}{\begin{equation}}
\newcommand{\bel}[1]{\begin{equation}\label{#1}}
\newcommand{\ee}{\end{equation}}
\usepackage{xcolor}

\newcommand{\Dan}[1]{{\color{red}\bf [Dan: #1]}}

\newcommand{\ire}[1]{{\color{green} \bf [Irene: #1]}}

\begin{document}
\title{
Decoding Quantum Field Theory with Machine Learning}
\author{Daniel Grimmer}
\email{daniel.grimmer@philosophy.ox.ac.uk}
\affiliation{Reuben College, University of Oxford, Oxford, OX1 1DW, UK}
\affiliation{Faculty of Philosophy, University of Oxford, Oxford, OX2 6GG United Kingdom}

\author{Irene Melgarejo-Lermas}
\email{i2melgar@uwaterloo.ca}
\affiliation{Institute for Quantum Computing, University of Waterloo, Waterloo, Ontario N2L 3G1, Canada}
\affiliation{Department of Applied Mathematics, University of Waterloo, Waterloo, Ontario N2L 3G1, Canada}

\author{Jos\'{e} Polo-G\'{o}mez}
\email{jpologomez@uwaterloo.ca}
\affiliation{Institute for Quantum Computing, University of Waterloo, Waterloo, Ontario N2L 3G1, Canada}
\affiliation{Department of Applied Mathematics, University of Waterloo, Waterloo, Ontario N2L 3G1, Canada}
\affiliation{Perimeter Institute for Theoretical Physics, Waterloo, Ontario N2L 2Y5, Canada}

\author{Eduardo Mart\'{i}n-Mart\'{i}nez}
\email{emartinmartinez@uwaterloo.ca}
\affiliation{Institute for Quantum Computing, University of Waterloo, Waterloo, Ontario N2L 3G1, Canada}
\affiliation{Department of Applied Mathematics, University of Waterloo, Waterloo, Ontario N2L 3G1, Canada}
\affiliation{Perimeter Institute for Theoretical Physics, Waterloo, Ontario N2L 2Y5, Canada}

\begin{abstract}
We demonstrate how one can use machine learning techniques to bypass the technical difficulties of designing an experiment and translating its outcomes into concrete claims about fundamental features of quantum fields. In practice, all measurements of quantum fields are carried out through local probes. Despite measuring only a small portion of the field, such local measurements have the capacity to reveal many of the field's global features. This is because, when in equilibrium with their environments, quantum fields store global information locally, albeit in a scrambled way. We show that neural networks can be trained to unscramble this information from data generated from a very simple one-size-fits-all local measurement protocol. To illustrate this general claim we will consider three non-trivial features of the field as case studies: a) how, as long as the field is in a stationary state, a particle detector can learn about the field's boundary conditions even before signals have time to propagate from the boundary to the detector, b) how detectors can determine the temperature of the quantum field even without thermalizing with it, and c) how detectors can distinguish between Fock states and coherent states even when the first and second moments of all their quadrature operators match. Each of these examples uses the exact same simple fixed local measurement protocol and machine-learning ansatz successfully. This supports the claim that the framework proposed here can be applied to nearly any kind of local measurement on a quantum field to reveal nearly any of the field's global properties in a one-size-fits-all manner. 

\end{abstract}

\maketitle

\section{Introduction}\label{Introduction}

Our current best understanding of nature comes from quantum field theory (QFT). However, the process of obtaining experimental information through measurements from QFTs is arguably difficult to formalize. For example, projective measurements in QFT are incompatible with its relativistic nature:  they cannot be localized~\cite{Redhead1995}, they can introduce ill-defined operations~\cite{alvaro} and can enable superluminal signaling even in simple setups~\cite{sorkin,Dowker,Dowker2}. Nevertheless, from high-energy physics experiments at the LHC to the capture of light at the human retina, quantum fields are subject to measurements where data is extracted through their interaction with localized probes. Such probes (e.g., atoms being excited by the electromagnetic field) can be generally modeled by particle detectors~\cite{UnruhWald,genrel,Takagi}. Particle detectors perform indirect measurements on the field that are well-defined~\cite{Louko2006,Satz2007,Edu2015,TalesBruno2020,TalesBruno2021,Pipo2021,PipoMaria2023} and physically meaningful~\cite{Martin-Martinez2018,Lopp2020}, allowing us to formulate a consistent measurement theory for quantum fields~\cite{JoseLuisEdu,Hector2023Measurements}.

However, it is not always obvious how we are to translate outcomes of measurements performed on local probes into claims about concrete features of a quantum field. This is even more difficult if the measurements are local (as they are since realistic probes are necessarily localized systems) and we want to determine global properties of a field. Luckily for us, a thermalized quantum field stores information about its global structure locally, albeit in a very scrambled way ~\cite{donpage,edu16structure,Maldacena2016,yoshida2017efficient,Landsman2019,YAMAGUCHI20191255}. The question is then how sophisticated must our local measurement protocol (and the following data analysis) be in order to determine something of interest about the field from local measurement data. It is thinkable that if one is allowed sufficient measurements on a sufficiently large array of probes, one should be able to resolve any feature of interest of a quantum field, at least in principle. We say `in principle' because---except for very few simple cases---there is usually no direct way of translating the theoretical predictions of particle detectors into specific values for the targeted features or parameters of the field. That makes it even more challenging to translate the readouts of local detectors into concrete claims about the field that they measure. Typically, the best one can do when trying to distinguish different field configurations is to look at differences in the detectors' transition probabilities (see, e.g.,~\cite{Takagi,jorma2012,keith,AidaAndRobb,Smith_2014, keith, AidaDarkness}). 

Moreover, the probes we use to measure quantum fields are usually simple in nature, and certainly much simpler and with smaller Hilbert spaces than the QFT itself. Additionally, in most cases we will not  have access to a large array of probes with which to sample the field, but rather only a handful. Because of this, translating measurement data (e.g., a set of zeros and ones generated by measuring a two-level particle detector) into concrete claims about the field seems, a priori, a very complicated task.

In general, one may not expect a context-free solution to this question. One might expect that for each feature of the QFT we may be interested in, we will need to define 1) a different local measurement protocol and 2) a different data analysis procedure. This assumption, i.e., that in general one needs different experiments and data processing techniques to measure different things, is arguably accepted without questions in experimental physics. Our goal here is to show that this is not the case. Concretely, we will show that fixing a simple local measurement protocol and a basic machine-learning data analysis ansatz is enough to unravel a wide variety of non-local features of the QFT with a one-size-fits-all method.

Recently, machine learning has proven effective at processing data from quantum systems (see, e.g.,~\cite{ optAdaptMetro,cory,Carrasquilla2017,hui2018,Torlai2018,fedorov2019,SignProblem,Cimini2019,Lidiak2020,Carrasquilla2021,Genois2021,Luiz2022}) and QFTs in the lattice~\cite{PhysRevD.100.011501, PhysRevD.97.094506,LaDeBea,Chernodub2020,Boyda2021,Jiang2021,Shi2022}. We will show, beyond lattice theories, how the task of unscrambling non-local information from measurements that are local in space and time is one that machine learning is well suited for.

The core of this paper is to develop a universal framework to extract information about a QFT through local measurements. To demonstrate the breadth of this framework we will apply it to three examples. Namely, we will show how local probes can 1) learn about boundary conditions even before a signal has time to propagate from the boundary to the probe, 2) learn the KMS temperature of a field with great accuracy even before a probe has time to thermalize with the field, and 3) distinguish Fock and coherent states even when the first and second moments of all their quadrature operators match.


Our goal is to show that combining machine learning techniques with detector model tools from QFT allows us to avoid the complexity of designing specific local measurement protocols and translating their outcomes into concrete claims about the field. Rather, one can use a simple all-purpose measurement protocol. This offloads all the complexity to the data processing, which is handled via machine learning. The success of this framework in the three substantially different case studies presented should also be strong evidence of its generality: it can be applied regardless of the specific targeted feature of the QFT, hence in a universal, context-free manner.



\vspace{-0.05cm}

\section{General Framework}\label{GeneralFramework}

In this Section we will introduce the general measurement protocol and data processing framework that constitute the core of our proposal. To make things concrete, let us assume that we have some foliation of spacetime associated to the coordinate frame $(t,\bm{x})$ and a probe system, D, coupled to a quantum field, $\hat{\phi}(t,\bm{x})$, in a local way. In particular, we will take the probe to be coupled linearly to the field via one of its observables $\hat{\mu}_\textsc{d}(t)$ in the interaction picture\footnote{This prescription can actually be done explicitly in a covariant manner at the level of Hamiltonian densities. See, e.g.,~\cite{TalesBruno2020,TalesBruno2021} for details.} as 
\begin{align}\label{Hint0}
\hat{\mathcal{H}}_\textsc{int} = \lambda \, \chi(t) \int_{\mathbb{R}^n} \!\!\!\!\d^n \bm{x} \, F(\bm{x}) \, \hat{\mu}_\textsc{d}(t) \otimes \hat{\phi}(t,\bm{x}),
\end{align}
where the switching function $\chi(t)$ and the smearing $F(\bm{x})$ characterize the locality of the interaction in time and space respectively. This coupling is motivated by the Unruh-DeWitt model~\cite{unruh1976,genrel}, which captures the fundamental features of the light-matter interaction when exchange of angular momentum can be neglected~\cite{Martin-Martinez2018,Edu2013,pozashyd,Lopp2020}. 

Imagine that we are interested in some global property of the field, for instance we might want to know whether this field lives in a spacetime with an open topology versus a closed one. To determine this, we might carefully design a measurement protocol, $M(\textsc{top})$, where we let a probe (or an array of them) interact with the field in a localized manner, and then perform measurements on the probe (or array of probes) to extract some data $D\in\mathbb{R}^{N_m}$ where $N_m$ is the number of measurements performed on the probe. A very simple example would be if $M(\textsc{top})$ specifies that we are to couple one single probe to the field for 1 second and then measure $\hat\mu_\textsc{d}$, repeating this process $N_m$ times.

In addition to specifying the data collection process, we must also design a data-analysis function, i.e., \mbox{$f_{\textsc{top},M(\textsc{top})}:D\to\{\text{open},\text{closed}\}$}, which will (hopefully) determine with high accuracy the topology of spacetime from the data. Here, $f_{\textsc{x},M}$ is a data analysis function which produces claims about $\textsc{X}$ from data generated by the measurement procedure $M$. Simply put, given a feature of interest $\textsc{X}$, $M(\textsc{x})$ is the measurement scheme designed by the experimentalist to inform us about X specifically, and $f_{\textsc{x},M(\textsc{x})}$ is the ``dictionary'' that translates the experimental data collected during the measurement into explicit statements about $\textsc{X}$, with some degree of accuracy. 

The problem now is how we might go about designing a good $M(\textsc{top})$ and $f_{\textsc{top},M(\textsc{top})}$. In designing $M(\textsc{top})$ one might try to find measurements which are particularly well suited for the identification of the topology. In particular, one might wish to produce data which is, somehow, explicitly revealing of the field's topology. In other words, if the experimenter does a good job designing $M(\textsc{top})$, then $f_{\textsc{top},M(\textsc{top})}$ could be relatively trivial, i.e. the data would require little analysis. Unfortunately, this approach is not generally available to us. QFT forces us to use local measurements, and as a consequence some non-trivial data analysis is required to piece these local probe measurements together into a global picture that tells us about topology~\cite{edu16structure}. 

However, even if we could find a suitable $M(\textsc{top})$ and $f_{\textsc{top},M(\textsc{top})}$ to determine the topology of a spacetime through the measurement of a quantum field, their utility would be limited to this particular feature. If we are interested in a different feature, $\textrm{F}$ (the charge of the field, its mass, entanglement structure, the space-time geometry, etc.), we would likely need to design a very different measurement protocol, \mbox{$M(\textsc{f})$}---which is well-suited to $\textrm{F}$---and a new data analysis function, \mbox{$f_{\textsc{f},M(\textsc{f})}$}. While we may not have to design \mbox{$M(\textsc{f})$} and  \mbox{$f_{\textsc{f},M(\textsc{f})}$} from scratch---many intuitions and previous knowledge may be applicable---this redesign is likely to be non-trivial. This seemingly uncontroversial statement, that you need different experiments and data processing techniques to measure different things, is one of the central (often unquestioned) tenets of experimental design in physics. 

In this light, a question we address in this paper is whether when we shift our interest to feature $\textsc{F}$ we can be lazy and keep our old measurement protocol, $M(\textsc{top})$. This would completely transfer the burden from designing a measurement protocol to designing a data-analysis function. In particular, we would now have to construct a function, \mbox{$f_{\textsc{f},M(\textsc{top})}$}, which extracts information about $\textrm{F}$ from data produced from $M(\textsc{top})$. 

One may think this lazy strategy will not work for two main reasons. Firstly, one may have the intuition that since $M(\textsc{top})$ was not designed with feature $\textsc{F}$ in mind, finding a good \mbox{$f_{\textsc{f},M(\textsc{top})}$} is likely to be difficult, if not impossible. Indeed, if we then changed our interest to some other feature $\textsc{F}'$ we would face the same difficult (if not impossible) task of finding a good \mbox{$f_{\textsc{f}',M(\textsc{top})}$}. Secondly, one may be inclined to think that even if $M(\textsc{top})$ somehow does extract a sufficient amount of information to be able to discern from it a wide variety of features, 
then it is likely to be an excessively complicated measurement procedure.

However, as we will show in this paper, both of these concerns can be  overcome. In particular, we will present a simple fixed local measurement protocol, $M_0$, designed without any specific feature of the QFT in mind. Despite this simplicity, we will show that the data produced by $M_0$ can be processed to produce accurate conclusions about a wide variety of features of the QFT. Moreover, we will prove that the data-analysis functions $f_{\textsc{f},M_0}$ which produce these conclusions can be easily generated from a basic machine-learning ansatz using standard supervised learning techniques.

This also connects with a fundamental question in the measurement theory of QFT. How do quantum fields store information? The fact that untargeted local measurements can be used to track non-local features of a QFT can be ultimately traced back to the fact that quantum fields tend to store global information locally (albeit perhaps in a scrambled way). Hence, simple local measurements are sufficient to extract global information, and machine learning is an effective way of unscrambling it; Indeed, extracting non-trivial features from complex data sets is exactly the kind of task that machine learning is well-suited for. As we will see, the ability of neural networks to unscramble global features of the field from local probe information informs us about how fields store information as they react to changes in their environment.

\subsection{Simple Fixed Measurement Procedure}
In this Subsection we propose a simple measurement protocol (what we denoted $M_0$ above) to produce labeled data from a probe coupled locally to a quantum field. This data that can then be processed to learn about different features of QFT.

In order to make things concrete, we will now specify some details about the probe and its interaction with the field. It is critical to note that the methods discussed in this paper are not dependent on the particular details of the field, the probe or their interaction, but we will particularize the choice of probe to the usual Unruh-DeWitt model in Eq.~\eqref{Hint0}, a common simple (yet realistic~\cite{Martin-Martinez2018,pozashyd,Edu2013,Lopp2020}) model for particle detectors probing quantum fields.

Consider a local probe coupled to a quantum field linearly via~\eqref{Hint0}. For illustration purposes, we model the probe as a harmonic oscillator with free Hamiltonian \mbox{$\hat{\mathcal{H}}_{\textsc{d}} 
=\hbar\omega_\textsc{d}(\hat{p}_\textsc{d}^2+\hat{q}_\textsc{d}^2)/2$}, where $\hat{q}_\textsc{d}$ and $\hat{p}_\textsc{d}$ are the probe's unitless quadrature operators satisfying \mbox{$[\hat{q}_\textsc{d},\hat{p}_\textsc{d}]=\ii\,\hat{\openone}$}. We take the probe to couple to the field via $\hat{\mu}_\textsc{d}=\hat{q}_\textsc{d}$. Our measurement procedure is as follows:
\begin{enumerate}
\item Initialize the field according to some conditions labeled $y$. For instance, for a quantum field in an optical cavity, if $y$ labels boundary conditions, this would mean preparing a cavity with those boundary conditions and letting the field equilibrate with the cavity walls. If $y$ labels temperatures, this would mean preparing a cavity at that temperature and letting the field thermalize with the cavity.
\item Initialize the probe to its ground state. Couple the probe locally to the field at time $t=0$ (i.e., the switching function $\chi(t)$ is zero before $t=0$).
\item At time $t_\textsc{m}=T_\textsc{min}>0$, perform a projective measurement on the probe's $\hat{q}_\textsc{d}$ quadrature and record the result. 
\item Repeat steps $1$ to $3$ but measuring the probe's $\hat{p}_\textsc{d}$ quadrature. Then repeat steps $1$ to $3$ but measuring the probe's \mbox{$\hat{r}_\textsc{d}=(\hat{q}_\textsc{d}+\hat{p}_\textsc{d})/\sqrt{2}$} quadrature.
\item  Repeat steps $1$ to $4$ a total of $N_\textsc{times}-1$ more times increasing $t_\textsc{m}$ by an amount $\Delta t$ each time.
\item Repeat this whole process $N_\textsc{tom}$ times.
\end{enumerate}
This (simple and untargeted) measurement procedure yields raw data $D_\textsc{raw}\in\mathbb{R}^{N_m}$, where \mbox{$N_m=3\times N_\textsc{times}\times N_\textsc{tom}$}. The goal of the $N_\textsc{tom}$ repetitions is to increase the precision with which we can calculate the averages $\langle \hat q_\textsc{d} \rangle$, $\langle \hat p_\textsc{d} \rangle$, and $\langle \hat r_\textsc{d} \rangle$. That is, the higher $N_\textsc{tom}$, the more accurate the state tomography we can perform on the detector. This data comes along with an associated label, $y$. We collect $N_\textsc{samples}$ of these pairs, $(D_\textsc{raw},y)$, where each of these $N_\textsc{samples}$ data points $D_\textsc{raw}$ are associated with generally different labels $y$.


As we will see, this simple local (in both time and space) interaction of the detector with the field produces enough information to determine a variety of non-local properties of the field. To accomplish this, we will use this labelled data to train a neural network. Once trained the neural network will be able to accurately predict the correct label $y$ when given new unlabeled data. In this way we will be able to learn about non-local features of the field from our local measurement data and the trained network. Some non-local features of the QFT might be more effectively captured by using more than one detector, so that we can use not only the outcomes of their individual measurements but also the correlations between them. In this sense, note that the above measurement scheme can be straightforwardly generalized to use \textit{arrays} of local probes.

As a final remark, it is worth noticing that it is not necessary that the training (and validation) data come from actual experiments as described above. Indeed, it will often be convenient to train the neural network using simulated data from the available theoretical models to prepare the network to identify features in experimental data.



\vspace{-0.7cm}

\subsection{Data Preprocessing}
While we could train our neural network directly on our
$N_\textsc{samples}$ labeled data points, $(D_\textsc{raw},y)$, in order to improve and speed up the training we will first compress and preprocess the data.

Note that each of our $N_\textsc{tom}$ measurements of $\hat{q}_\textsc{d}(t_\textsc{m})$ where \mbox{$t_\textsc{m}=T_\textsc{min}+\textsc{m}\,\Delta t, \ \textsc{m}\in\{0,\dots,N_\textsc{times}-1\}$} are independent and identically distributed; they each come from identical independent experiments. We can summarize these $N_\textsc{tom}$ measurement outcomes, $q_k$, via their sample mean, sample variance, and sample fourth central moments,
\begin{align}
\label{samplemean}
\bar{q}(t_\textsc{m})
&=\frac{1}{N_\textsc{tom}}\sum_{k=1}^{N_\textsc{tom}} q_k\\    
\bar{s}_{q}(t_\textsc{m}) 
\label{samplesecond}
&= \frac{1}{N_\textsc{tom}}\sum_{k=1}^{N_\textsc{tom}} \big(q_k-\bar{q}(t_\textsc{m})\big)^2\\
\label{samplefourth}
\bar{s}_{4,q}(t_\textsc{m})
&=\frac{1}{N_\textsc{tom}}\sum_{k=1}^{N_\textsc{tom}} (q_k-\bar{q}(t_\textsc{m}))^4
\end{align}
We can similarly compress our measurements of $\hat{q}_\textsc{d}$, $\hat{r}_\textsc{d}$ and $\hat{p}_\textsc{d}$ at each time $t_\textsc{m}=T_\textsc{min}+\textsc{m}\Delta t$. Depending on our needs, we might want to include higher order (e.g., eighth) sample central moments. However, we will see that including the sample fourth moments (and often even only the second moments) in our compressed data is enough to allow our proposed machine learning methods to address our three physical examples, and potentially to answer a very general breadth of questions about the quantum field. 

Once compressed, our data is described by the time series
\begin{align}
\Big\{&\bar{q}(t_\textsc{m}),& &\bar{r}( t_\textsc{m}),& &\bar{p}(t_\textsc{m}),\\
\nonumber
&\bar{s}_{q}(t_\textsc{m}),& &\bar{s}_{r}( t_\textsc{m}),& &\bar{s}_{p}(t_\textsc{m}),\\
\nonumber
&\bar{s}_{4,q}(t_\textsc{m}),& &\bar{s}_{4,r}( t_\textsc{m}),& &\bar{s}_{4,p}(t_\textsc{m})\Big\}_{\textsc{m}=0}^{N_\textsc{times}-1}.
\end{align}
That is, our compressed data can be represented by a vector \mbox{$D_\textsc{c}\in\mathbb{R}^d$} where $d=9\,N_\textsc{times}$. Notice that even after compression the data will generally be high-dimensional since $N_\textsc{times}$ will be very large.


After compression we perform standard preprocessing~\cite{Goodfellow}: we center the data, do principal component analysis, and whiten the data. The details of our preprocessing are discussed in Appendix \ref{appendixA}. Let us call the preprocessed data, $D_\textsc{p}$. Next we will discuss how our neural network is trained and validated on the $N_\textsc{samples}$ labeled data points $(D_\textsc{p},y)$.

\subsection{Neural Network Training}
In this Section we lay out how neural networks provide a data-analysis ansatz ($f_{\textsc{f},M_0}$ in Sec.~\ref{GeneralFramework}) which we can use to process our data. In particular, we will discuss how we can analyze the data produced by our simple fixed local measurement protocol, $M_0$, to arrive at accurate conclusions about a wide variety of (non-local) features of the QFT. 

\begin{figure}
\includegraphics[width=0.48\textwidth]{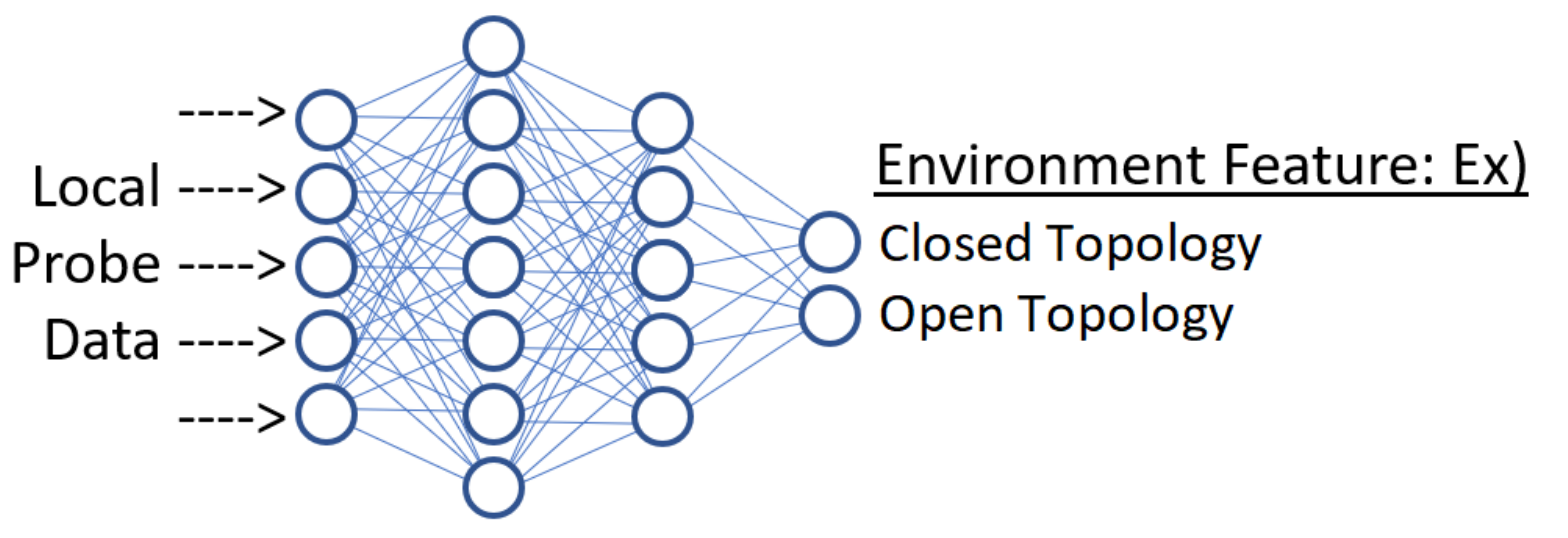}
\caption{(Color online.)  A schematic example of a neural network for processing local probe data to learn about a global feature of a QFT.}\label{NN}
\end{figure}

Neural networks model complicated high-dimensional functions by alternatively applying tunable linear-affine transformations (controlled by weights and biases) and fixed non-linear transformations (i.e., activation functions) to their inputs. Fig. \ref{NN} illustrates a simple neural network architecture one might use for classifying features of interest (in our example, the topology of the spacetime) based on local probe measurement data on a quantum field. The circles in Fig.~\ref{NN} represent the fixed activation functions and the lines represent the tunable weights and biases. The neural network takes as input the local probe measurement data and outputs a probability assignment for each possible value of the feature of interest (whether the spacetime topology is open or closed).

In order to find the proper settings for the network's weights and biases (in the example in Fig.~\ref{NN}, those settings which accurately predict the spacetime topology) we follow a supervised training procedure. Supervised training requires labeled data, i.e., many datapoints, $D_\textsc{p}$, each paired with a label, $y$, indicating the result that our data analysis should produce. In the topology example this is $y\in\{\text{open},\text{closed}\}$. These labels may also be continuous, for instance if we were interested in determining, e.g., the field's mass, $m$, we would have \mbox{$y=m\in\mathbb{R}_+$}.

Given $N_\textsc{samples}$ labelled datapoints we divide these into data for training the neural network (\mbox{$N_\textsc{train}=0.75\,N_\textsc{samples}$}) and for validating its accuracy (\mbox{$N_\textsc{valid}=0.25\,N_\textsc{samples}$}). We then define a cost function to characterize how wrong our network's predictions are over the training data. The network's weights and biases are adjusted to minimize this cost function. Once the training is complete, the network's accuracy is evaluated on the validation data.

The network architecture (fully connected feedforward) and training procedure (stochastic gradient descent) that we use are standard~\cite{Goodfellow}. Our code can be found on GitHub~\cite{github}, and a summary of our network architecture, data preprocessing, and training process is given in Appendix \ref{appendixA}.

\section{Two Methods of Simulating Data Generation}\label{TwoMethods}
In the previous Section we  discussed how the probe will interact with the field, how it will be measured, how the resulting data will be processed, how this data will be used to train a neural network, and how the accuracy of this network will be validated. In Sections~\ref{BoundarySensing}, \ref{Thermometry}, and \ref{FockExample} we will apply this process to three examples: remote boundary sensing, thermometry, and quantum state discrimination, respectively. To be able to do this we need to first discuss  how exactly we produce the labeled data in a general QFT framework. This is what we do in this Section.

Since we are but poor theoreticians, rather than having a real-life probe interact with a real-life field to generate our training data, we instead simulate both the probe-field interaction and the probe's subsequent measurement. Notice however that training and validating with simulated data is not necessary: If one were to work with actual gathered experimental data the exact same analysis would apply. The methods discussed in this paper are independent of how the data is generated, be it by experiment or by simulation. Moreover, our methods are independent of exactly how we simulate the data generation process. To demonstrate this we will simulate the probe's response in two different ways: one involving a lattice approximation and  one involving Dirac-delta interactions between the probe and the field in a continuous optical cavity. In both cases we will take the interaction Hamiltonian between the probe and the field as a realistic and accepted model for the light-matter interaction of atoms with the electromagnetic fields~\cite{Martin-Martinez2018,pozashyd,Edu2013,Lopp2020}.

Although the methods discussed in this paper are not dependent in any way on the particular details of the field,  for concreteness and as a proof-of-principle, we study a probe coupled to a 1+1D massive scalar field as in \eqref{Hint0} with $\hat{\mu}_\textsc{d}=\hat{q}_\textsc{d}$ (see e.g.,~\cite{unruhzurek1989,humatack1994,MassarSpindel2006, ShihYuinHu2007, EduIvetteRob2011,IvetteBruschiLee,cavitiesEdu}).  The field's free Hamiltonian is
\begin{align}\label{Hamil1} 
\hat{\mathcal{H}}_{\phi} 
&=\dfrac{1}{2}\int \!\d x \!\left(\!c^2 \hat \pi(t,x)^2 \!+\! (\partial_x\hat\phi(t,x))^2 \!+\! \dfrac{m^2\!c^2}{\hbar^2} \hat \phi(t,x)^2\!\right)\!,
\end{align}
where $\hat{\phi}(t,x)$ and $\hat{\pi}(t,x)$ are field observables satisfying the canonical commutation relations \mbox{$[\hat{\phi}(t,x),\hat{\pi}(t,y)]=\ii\hbar\,\delta(x-y)\hat{\openone}$}.  

\subsection{Method 1: Lattice Approximation}
\label{method1}
For two of the three examples that we will analyze (boundary sensing and thermometry) we simulated the probe's response to the field using a lattice approximation. This approximation is not necessary but it does greatly simplify the analysis. It can be motivated as follows.

From \eqref{Hint0}, if the region where $F(x)$ is non-negligible has a lengthscale $\sigma$, the probe will not couple to field modes with wavenumber $\vert k\vert\gg \sigma^{-1}$~\cite{Emma}. Taking the probe to have a Gaussian smearing with standard deviation $\sigma$, the coupling to the high-frequency field modes is exponentially suppressed. This motivates a UV-cutoff~\cite{RevModPhys.51.659, Gattringer:2010zz} at \mbox{$\vert k\vert \leq K\coloneqq16/\sigma$} (see Appendix~\ref{appendixB} for technical details).

Cutting these high-frequency modes out of both the interaction Hamiltonian \eqref{Hint0} and the field's free Hamiltonian \eqref{Hamil1} strongly simplifies the situation without substantially affecting the probe's response to the field. Typically, UV-cutoffs are at odds with locality assumptions, as they break the causality of the theory \cite{Edu2015}. However, for our choice of scales, the UV-cutoff only introduces a relative error in the probe's response below a fraction of a percent (see Appendix~\ref{appendixB} for details) and therefore the field theory remains effectively causal to any perceptible accuracy. The causal behaviour of this scenario is explicitly demonstrated in our first example (see Sec.~\ref{BoundarySensing}).

Taking the UV-cutoff can be seen as placing a bandlimit on the field. The Nyquist--Shannon sampling theorem then allows us to exactly reconstruct this bandlimited field from a discrete lattice of sample points. We provide some discussion in Appendix~\ref{appendixB}. This yields the Hamiltonians
\begin{align}\label{HPhiUV}
    \hat{\mathcal{H}}_{\phi}^\textsc{uv} 
    &= \sum_n \dfrac{m c^2}{2}
    \left(\hat{p}_n^2 + \hat{q}_n^2\right) +\dfrac{\hbar^2}{2 m a^2}(\hat{q}_{n+1} - \hat{q}_n)^2,\\
    \label{HIntUV}
    \hat{\mathcal{H}}_\textsc{int}^\textsc{uv} &= \lambda_0 \, \chi(t) \, \sum_n a\,F(x_n) \,  \hat{q}_\textsc{d}\otimes \hat{q}_n,
\end{align}
where $a=\pi/K$ is the lattice spacing , \mbox{$x_n\coloneqq n\,a$} and $\lambda_0=\lambda \, \hbar/\sqrt{a\,m}$ is strength of the probe-field coupling. We have defined dimensionless field operators \mbox{$\hat{q}_n \coloneqq \sqrt{a m/\hbar^2} \, \hat{\phi}(t,x_n)$} and 
\mbox{$\hat{p}_n \coloneqq \sqrt{a/m} \, \hat{\pi}(t,x_n)$} satisfying the canonical commutation relations \mbox{$[\hat{q}_i,\hat{p}_j]=\ii\,\delta_{ij}\,\openone$}. Since the operators $\hat{q}_n$ and $\hat{p}_n$ are associated with the field operators in a localized region we say that $(\hat{q}_n,\hat{p}_n)$ define \textit{spatial modes}.

It is also possible to restrict the field theory to a finite region of space (think of an optical cavity or a finite size transmission line). We can incorporate this to our general formalism, which results in an additional IR-cutoff after we restrict the field to the region \mbox{$x\in[0,L]$} where \mbox{$L=N\,a$}.

Given the above Hamiltonian, we can use Gaussian Quantum Mechanics~\cite{Simon1988,Ferraro2005,GQMRev} to efficiently simulate the dynamics. This is because the Hamiltonians controlling the dynamics are all quadratic in the field and probe quadrature operators,
\begin{align}\label{vectorofquadratures}
\hat{\bm{X}}
\coloneqq
(\hat{q}_\textsc{d},\hat{p}_\textsc{d},\hat{q}_1,\hat{p}_1,\hat{q}_2,\hat{p}_2,\dots)^\intercal \,,    
\end{align}
and because the initial states of the probe and field are Gaussian states (i.e., states with Gaussian Wigner functions\footnote{For an introduction to the phase space quantum mechanics formalism in arbitrary dimensions, see, e.g., Section 4.7 in~\cite{notesQT}.}). Together these guarantee that the probe and field states remain in Gaussian states throughout their evolution. As such, their quantum states are fully characterized by the first and second moments of $\hat{\bm{X}}$ which we can collect together in a displacement vector \begin{equation}\label{displacementvector}
\bm{X}=\langle \hat{\bm{X}} \rangle=(\langle\hat{q}_\textsc{d}\rangle,\langle\hat{p}_\textsc{d}\rangle,\langle\hat{q}_1\rangle,\langle\hat{p}_1\rangle,\langle\hat{q}_2\rangle,\langle\hat{p}_2\rangle,\dots)^\intercal   
\end{equation}
and a covariance matrix
\begin{equation}\label{covariancematrix}
\sigma_{ij}=\langle\{\hat{X}_i,\hat{X}_j\}\rangle/2-\langle \hat{X}_i\rangle \langle \hat{X}_j\rangle
\end{equation}
Unitary dynamics for the joint density matrix, \mbox{$\hat\rho\to \hat{U}\,\hat\rho\, \hat{U}^\dagger$} corresponds to symplectic(-affine) evolution in phase space. In particular, the changes in the displacement vector and the covariance matrix are given by \mbox{$\bm{X}\to S\,\bm{X}+\bm{d}$} and \mbox{$\sigma \to S\sigma S^\intercal$}, where $S$ is a symplectic matrix and $\bm{d}$ is a vector. Concretely,
\begin{align}
    S(t)= \mathcal{T}\!\exp \left(\frac{1}{\hbar}\int_{0}^{t}\d t' \ \Omega \mathcal{F}(t')\right),
\end{align}
where $\mathcal{F}(t)$ is the symmetric matrix that satisfies \mbox{$\hat{\mathcal{H}}=\hat{\mathcal{H}}_\textsc{d}+\hat{\mathcal{H}}_{\phi}+\hat{\mathcal{H}}_\textsc{int}(t)=\frac{1}{2}\hat{\bm{X}}^\intercal \mathcal{F}(t)\hat{\bm{X}}$}, $\Omega$ is the symplectic form, $[\hat{X}_j,\hat{X}_k]=\ii\,\Omega_{jk}\hat{\openone}$, and $\mathcal{T}$ is the time-ordering symbol. 
We note that since $\hat{\mathcal{H}}$ has no terms linear in $\bm{\hat{X}}$ we have $\bm{d}=0$. Moreover, we note that $S$ can be calculated non-perturbatively much more efficiently than $U$~\cite{Ferraro2005,AdessoThesis,notesQT,Adesso2014,GQMRev,lami,GaussianClass}.

Once we have computed the evolved joint covariance matrix and displacement vector, we can easily isolate the reduced state of the probe system as \mbox{$\bm{X}_\textsc{d}=(\mu_q,\mu_p)^\intercal=
(\langle\hat{q}_\textsc{d}\rangle, \langle\hat{p}_\textsc{d}\rangle)^\intercal$} and
\begin{align}
\sigma_\textsc{d}&= 
\begin{pmatrix}
\sigma_{qq} & \sigma_{qp}\\ 
\sigma_{pq} & \sigma_{pp}\\
\end{pmatrix} \\
&=\begin{pmatrix}
\langle q_\textsc{d}^2\rangle & \langle\{\hat{q}_\textsc{d},\hat{p}_\textsc{d}\}\rangle/2\\ \langle\{\hat{q}_\textsc{d},\hat{p}_\textsc{d}\}\rangle/2 & \langle p_\textsc{d}^2\rangle\\
\end{pmatrix}
-\bm{X}_\textsc{d}\bm{X}_\textsc{d}^\intercal \;. \nonumber
\end{align}
These values determine the distributions which our measurements of $\hat{q}_\textsc{d}$, $\hat{p}_\textsc{d}$ and $\hat{r}_\textsc{d}$ should be drawn from. Namely, the outcomes of the measurements are distributed as
\begin{align}\label{qBarr}
q_k
&\sim \mathcal{N}\left(\mu_q,\sigma_{qq}\right)\\
p_k
& \sim \mathcal{N}\left(\mu_p,\sigma_{pp}\right) \\
r_k
&\sim \mathcal{N}\left(\frac{\mu_q+\mu_p}{\sqrt{2}},\frac{\sigma_{qq}+2\sigma_{qp}+ \sigma_{pp}}{2}\right)
\end{align}
where $\mathcal{N}(\mu,\sigma)$ is the normal distribution with mean $\mu$ and variance $\sigma$. This can be straightforwardly justified from the fact that the partial Wigner function of the probe is still Gaussian and that its marginals correspond to the position and momentum distributions.

Moreover, from the evolved probe state and $N_\textsc{tom}$ we can determine the distributions from which the sample mean and sample variance are drawn. For instance,
\begin{align}\label{qBar}
\bar{q}
&= \frac{1}{N_\textsc{tom}}\sum_{k=1}^{N_\textsc{tom}} q_k
\sim \mathcal{N}\left(\mu_q,\frac{\sigma_{qq}}{N_\textsc{tom}}\right),\\
\label{s2bar}
\bar{s}_{q}
&= \frac{1}{N_\textsc{tom}} \sum_{k=1}^{N_\textsc{tom}} \big(q_k-\bar{q}\big)^2
\sim \sigma_{qq} \,  \frac{\chi^2(N_\textsc{tom}-1)}{N_\textsc{tom}},
\end{align}
where $\chi^2(k)$ is the chi-squared distribution with $k$ degrees of freedom. Note that while the sample means are distributed normally, the sample variances are not.

\subsection{Method 2: Non-Gaussian Wigner Functions}
\label{method2}
Although working with the Gaussian formalism when possible is very convenient for calculational purposes, the techniques discussed in the previous Subsection can be easily extended to drop the assumption that the field and probe are initially in Gaussian states. We will show this explicitly in our third example scenario in which we will attempt to distinguish between a coherent state (which is Gaussian) and a Fock state (which is not).

Even when the initial state's Wigner function $W(\bm{\xi})$ is non-Gaussian we can still understand the dynamics in terms of phase space evolution~\cite{notesQT}. A Gaussian unitary transformation $U_\text{G}$ (i.e., evolution generated by a quadratic Hamiltonian) still corresponds to a symplectic transformation for the vector of quadrature operators of the form described in Sec.~\ref{method1}, so that the dynamics induced by our unitary evolution operator, $\rho\to \hat{U}_\text{G}(t)\,\rho\, \hat{U}_\text{G}(t)^\dagger$  can be equivalently written in the Heisenberg picture as
\begin{align}
\hat{\bm{X}} 
\to \hat{U}_\text{G}(t)^\dagger
\,\hat{\bm{X}}\,\hat{U}_\text{G}(t)
=S(t)\,\hat{\bm{X}}+\bm{d}(t)\hat{\openone}
\end{align}
for some symplectic matrix $S(t)$ and some vector $\bm{d}(t)$. For our particular case, $\bm{d}(t) = 0$ since $\mathcal{H}$ does not have any terms linear in $\hat{\bm{X}}$. 

This transformation entails, in turn, a transformation in the states' Wigner function, independent of the Gaussianity of the states. Mathematically, this transformation can be written as
\begin{align}
W(\bm{\xi})\to W(S^{-1}(t)\bm{\xi}). 
\end{align}
We can use this to determine the final reduced probe state from the initial probe-field state by integrating over all of the field variables as 
\begin{align}
W_\textsc{d}(q_\textsc{d},p_\textsc{d};t)
&=\int \d\bm{\zeta} \, W(S^{-1}(t)(q_\textsc{d},p_\textsc{d},\bm{\zeta})) \,,
\end{align}
where $\bm{\zeta}=(q_1,p_1,q_2,p_2, \dots, )$ runs over all the field variables. Note that if the initial probe-field state is non-Gaussian then the final probe state will also be non-Gaussian. That is, we cannot characterize it by its first and second moments alone. In general, this final probe state will have non-trivial higher moments. 

However, the higher moments of the final probe distribution can be calculated straightforwardly from the higher moments of the initial probe-field distribution. 
We can use the central limit theorem to generate the sample means, sample variances and sample fourth central moments of Eqs. \eqref{samplemean}, \eqref{samplesecond} and \eqref{samplefourth}. For instance, from the second, fourth and eighth central moments of the marginal distribution of the $q$ quadrature of the probe,
\begin{align}
&\sigma_{qq} = \int \d p \,\d q \,(q-\mu)^2  \,W_\textsc{d}(q,p),\\
&\gamma_4 = \int \d p \,\d q \,(q-\mu)^4  \,W_\textsc{d}(q,p),\\
&\gamma_8 = \int \d p \,\d q \,(q-\mu)^8  \,W_\textsc{d}(q,p),
\end{align}
where
\begin{equation}
    \mu= \int \d p \,\d q \,q  \,W_\textsc{d}(q,p) \;,
\end{equation}
we have, for $N_\textsc{tom}$ sufficiently large,
\begin{align}\label{q1}
&\bar{q}\sim \mu+ \sqrt{\frac{\sigma}{N_\textsc{tom}}}\mathcal{N}(0,1),\\
\label{q2}
&\bar{s}_q
\sim \sigma + \sqrt{\frac{\gamma_4-\sigma^2}{N_\textsc{tom}}}\mathcal{N}(0,1),\\
\label{q4}
&\bar{s}_{4,q} \sim \gamma_4 + \sqrt{\frac{\gamma_8-\gamma_4^2}{N_\textsc{tom}}}\mathcal{N}(0,1).
\end{align}
In particular, if we consider states with $\langle \hat q_\textsc{d} \rangle=0$, as we will in the example in Sec.~\ref{FockExample}, then
\begin{align}\label{q1bis}
&\bar{q}\sim \sqrt{\frac{\langle \hat{q}_\textsc{d}^2\rangle}{N_\textsc{tom}}}\mathcal{N}(0,1),\\
\label{q2bis}
&\bar{s}_q
\sim \langle \hat{q}_\textsc{d}^2 \rangle + \sqrt{\frac{\langle \hat{q}_\textsc{d}^4\rangle-\langle \hat{q}_\textsc{d}^2\rangle^2}{N_\textsc{tom}}}\mathcal{N}(0,1),\\
\label{q4bis}
&\bar{s}_{4,q} \sim \langle \hat{q}_\textsc{d}^4\rangle + \sqrt{\frac{\langle \hat{q}_\textsc{d}^8 \rangle-\langle \hat{q}_\textsc{d}^4\rangle^2}{N_\textsc{tom}}}\mathcal{N}(0,1).
\end{align}
where
\begin{equation}
\langle \hat{q}_\textsc{d} ^n \rangle = \int \d p \,\d q \,q^n  \,W_\textsc{d}(q,p)
\end{equation}
is the $n$-th moment of the marginal distribution of the $q$ quadrature of the probe. 

\section{A first physical example: Remote Boundary Sensing}\label{BoundarySensing}
As a first application of our framework, we will use measurements of a local probe near one end of a cavity ($x \approx 0$) to learn about the location of the boundary at the other end of the cavity ($x\approx L$).

In this example we will employ the lattice approximation discussed in the previous section. That is, the field will be approximated as a finite chain of harmonic oscillators which we will call spatial modes. Note from Eq.~\eqref{HPhiUV} that these spatial modes have a nearest-neighbor coupling. As we will see, the lattice approximation does not significantly impact the relativistic compliance of our setup. Indeed, we will show that, in practice, signals do not propagate superluminally in our lattice.

To simulate different positions of the far boundary, we will modify the coupling in Eq.~\eqref{HPhiUV} so that we separate the coupling of the two oscillators furthest from the probe from the rest:
\begin{align}
\hat{\mathcal{H}}_{\phi}^\textsc{uv}=&\sum_{n=1}^{N} \dfrac{m c^2}{2}
    \left(\hat{p}_n^2 + \hat{q}_n^2\right) +\sum_{n=1}^{N-2}\dfrac{\hbar^2}{2 m a^2}(\hat{q}_{n+1} - \hat{q}_n)^2 \nonumber \\
    &+\dfrac{\hbar^2}{2 m a^2}\left(\hat{q}_{N}^2+\hat{q}_{N-1}^2 \right)+\hat{H}_\text{last} \;.
\end{align}
That is, if there are $N$ harmonic oscillators in the lattice, we will consider modifications to the coupling between oscillator $N$ and oscillator $N-1$. We summarize the modified couplings under consideration in Table~\ref{CaseTable}.
\begin{table}[ht]
    \centering
    \begin{tabular}{c|c|c|c}
    y-label & Name & $\hat{H}_\text{last}$ for $t<0$ & $\hat{H}_\text{last}$ for $t\geq0$\\
    \hline
    y\,=\,1 & Full Bond & $\ g \ \ \quad \hat{q}_{N-1}\otimes\hat{q}_{N}$ & Same as $t<0$\\
    y\,=\,2 & Cut Bond & $\ 0 \ \ \quad \hat{q}_{N-1}\otimes\hat{q}_{N}$ & Same as $t<0$\\
    y\,=\,3 & Signal & $\ 0 \ \ \quad \hat{q}_{N-1}\otimes\hat{q}_{N}$ & $\ g \quad \hat{q}_{N-1}\otimes\hat{q}_{N}$\\
    \end{tabular}
    \caption{Modifications of $\hat{H}_\text{last}$ connecting the last spatial mode to the rest of the lattice. $g=\hbar^2/m a^2$ is the site-to-site coupling strength.}\label{CaseTable}
\end{table}

For the first case ($y=1$) we take $\hat{H}_\text{last}$ to be just the same as every other site-to-site coupling, so that the boundary is at the last site of the lattice. In the second case ($y=2$) we take $\hat{H}_\text{last}=0$, thus setting the boundary at the second to last site. We added a third case ($y=3$) to measure the time that it takes for the probe to detect a perturbation coming from the boundary. To do so, we consider a time-dependent coupling between the two last oscillators which turns on at $t=0$.
\begin{figure}[t]
\includegraphics[width=0.48\textwidth]{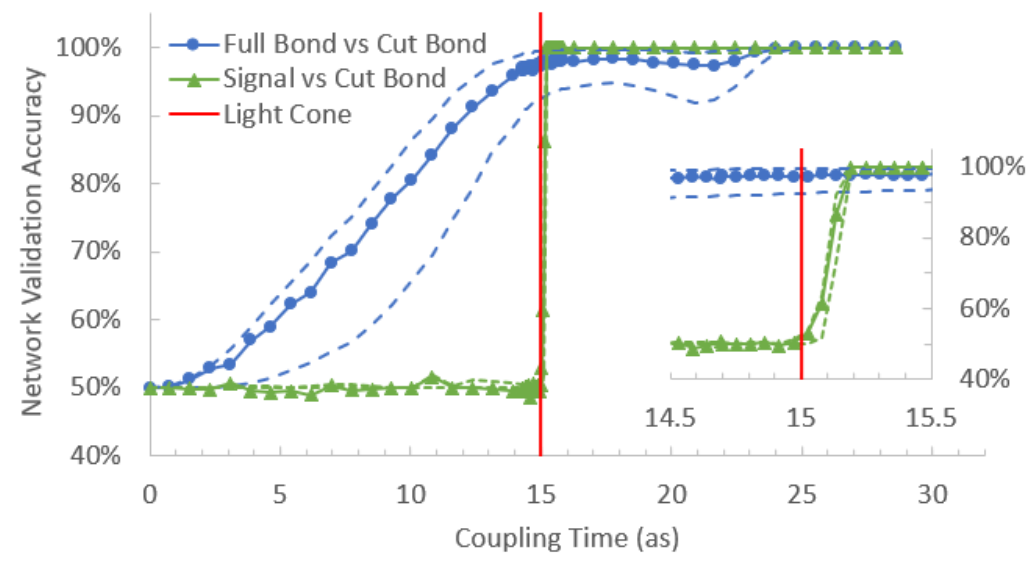}
\caption{We trained a neural network to predict (through classification) the position of the boundary of a quantum field in a cavity from local probe data gathered far from the boundary. The network was asked 1) to detect a signal sent from the boundary (green triangles) and 2) to detect a modification of the field's boundary position (blue circles). The network's accuracy (solid) along with upper and lower bounds on the theoretical optimal accuracy (dashed) are plotted as a function of the duration of the probe's interaction with the field. A point plotted at time $t$ indicates the network's accuracy given measurements taken at  $N_\textsc{times}=10$ measurement times between $t$ and the previous plot point. The network was trained on $N_\text{train}=11250$ examples. Each example summarizes $N_\textsc{tom}=10^{22}$ measurements of each of the probe's quadratures ($\hat{q}_\textsc{d}$, $\hat{r}_\textsc{d}$ and $\hat{p}_\textsc{d}$) at each measurement times. The inset shows details of the causal response of the detector to the signal. The vertical red line is at the edge of detector to boundary light crossing time.}\label{Results1}
\end{figure}

In each of these three cases we assume that the field has thermalized to the ground state of its $t<0$ Hamiltonian well before $t=0$. Recall that $t=0$ marks the instant when the probe first couples to the field. In this example, we take the switching function $\chi(t)$ to be constant for $t>0$. 

In the first two cases the $t\geq0$ field Hamiltonian is exactly the same as the $t<0$ Hamiltonian. All that changes at $t=0$ in these cases is that the probe couples to the field at one end of the cavity, $x\approx0$. We thus expect disturbances to begin propagating away from $x=0$ beginning at $t=0$. In the third case the field Hamiltonian suddenly changes at $t=0$. This change is localized around $x\approx L$. In this case we thus expect disturbances to begin propagating away from both $x=0$ and $x=L$ beginning at $t=0$. In this third case, we take the last spatial mode to be in a highly-squeezed state (8 dB~\cite{Vahlbruch2016}) before $t=0$ so that it produces a notable disturbance when it couples. 

By comparing cases 2 and 3 we can explicitly measure the signal-propagation speed on the lattice. In these cases the field is in exactly the same state prior to $t=0$, they differ by the local disturbance at $t=0$ and $x\approx L$ in case 3. For $t>0$ the disturbance at $x\approx L$ begins propagating through the lattice towards the probe. We can define the effective signaling time as the time it takes the probe to differentiate between cases 2 and 3. If the probe is able to differentiate cases 1 and 2 in less than this signalling time it cannot be due to having received a signal from the boundary. We can compare this signalling time to the light-crossing time of the cavity ($t=L/c$) to see if our probe is receiving any faster-than-light signals due to the  approximations employed.

We consider a detector of roughly atomic size, with a Gaussian smearing function of width $\sigma=53\text{ pm}$ (the Bohr radius). Taking the UV-cutoff $K=16/\sigma$ gives us a lattice spacing $a=\pi/K=10.4\text{ pm}$. We take the boundary to be at \mbox{$L=90\,\sigma=457\,a=4.7\text{ nm}$}. We set the detector's excitation energy $\hbar\omega_\textsc{d}=130\text{eV}$ and the field's mass \mbox{$m c^2=1\text{eV}$}. The field mass is much smaller than any other energy scale in the problem (effectively massless). Finally, we investigate the non-perturbative strong-coupling regime where \mbox{$\lambda_0=\hbar\omega_\textsc{d}=130\text{ eV}$}. Note that the choice of parameters is for demonstration purposes; similar results are also obtained for a large set of different parameters.

In Fig.~\ref{Results1} we show the performance of the neural network (solid line). The green triangle lines show the causal behaviour of the setup: when we send a signal from the far boundary to the detector (coupling a new oscillator at \mbox{$t=0$}) the neural network accuracy indicates that the probe does not receive the signal before \mbox{$\approx15\text{ as}$}. A comparison between this and a conservative estimate of the signal-to-edge-of-detector light-crossing time, \mbox{$(L-5\sigma)/c=15\text{ as}$} (vertical red line), shows that our toy model displays a good causal behaviour. This confirms that the lattice approximation did not compromise the relativistic structure that is essential for the model to be a faithful simulation of a quantum field theory in a cavity. 

The blue circle line represents the ability of the neural network to sense the boundary at the far end of the cavity. Here, the information about the boundary has had time to spread all over space in the equilibration process before $t=0$: the ground state knows locally about its boundary conditions. Indeed, the network accuracy shows that the nature of the field boundary can be resolved long before any signal from the boundary propagates to the detector. This allows the probe to see the boundary `without light', that is, in the vacuum state of the theory and much before the light-crossing time of the lattice. 

Notice that we do observe an increase in the accuracy of the neural network as the time of the measurements is further away from the start of the coupling of the probe. This is not related to signalling from the boundary, as the comparison between cases 2 and 3 shows, but rather to the fact that the more time we let the detector interact with the field, the more knowledge it gathers from the infrared structure of the field state, where the information about the boundary lies. 

In itself, this `seeing without light' phenomena is not a new result. It has been seen and understood in a number of different contexts~\cite{keith,Smith_2014,AidaAndRobb,AidaDarkness}. What is new here is the explicit collection of (simulated) data and the direct translation of these local measurement results into claims about the field's boundary conditions. Remarkably, the local measurement protocol and data-analysis ansatz was not tailored to the detection of boundaries. In fact (as we will soon see) the exact same local measurement protocol and data-analysis ansatz can be used to make accurate claims about a wide range of other features of the QFT, and not only to ``see without light''. 


\subsection{Near Optimality of the Neural Network}

Unlike the more complex scenarios in the sections below, this particular example admits a more conventional statistical treatment that will allow us to discuss in some detail how we know that the neural network in this example is behaving near-optimally.

Recall from~\eqref{qBar} and~\eqref{s2bar} that we know the distributions from which the sample first and second moments are drawn from. Thus, ultimately, we know the distributions which our compressed data $D_\textsc{c}$ is drawn from. In particular we know the distributions $p(D_\textsc{c}\vert y=1,\theta)$, $p(D_\textsc{c}\vert y=2,\theta)$, and $p(D_\textsc{c}\vert y=3,\theta)$ for each of the three cases listed in Table \ref{CaseTable}, where $\theta$ contains all the other parameters of the measurement setup (e.g., coupling times). These distributions are simple, yet made out of rather unwieldy combinations of Gaussians and $\chi^2$ distributions. 

Suppose that we are presented with some compressed data $D_\textsc{c}$ and asked to guess whether it came from a $y=1$ or a $y=2$ case. This is, in fact, the exact question we are repeatedly asking our neural network. For this binary classification task, the optimal solution is known: we ought to guess whichever $y$ makes $p(D_\textsc{c}\vert y,\theta)$ larger (see Appendix~\ref{appendixC} for details). The success rate of this strategy is $p_\text{success}(\theta)=\frac{1}{2}(1+\text{TV}_{12}(\theta))$ where $\text{TV}_{12}(\theta)\in[0,1]$ is the total variation distance between \mbox{$p(D_\textsc{c}\vert y=1,\theta)$} and \mbox{$p(D_\textsc{c}\vert y=2,\theta)$}. 

If we could calculate $\text{TV}_{12}(\theta)$ then we would have a tight upper bound on the accuracy achievable by any method aimed to distinguish these two cases. In particular, we would have an upper bound on the validation accuracy achievable by any neural network (with any architecture, training time, training method, etc.). Unfortunately, for the rather cumbersome combination of Gaussians and $\chi^2$ distributions pertaining to this example, $\text{TV}$ is not calculable in closed form. However, in the large $N_\textsc{tom}$ regime these distributions each simplify to multi-variate Gaussian distributions. While the total variation distance between multi-variate Gaussian distributions is unknown in general, we can compute the Hellinger distance $H(\theta)\in[0,1]$ between them. The Hellinger distance bounds the total variation distance above and below as (see Appendix~\ref{appendixC} for details)
\be
H(\theta)
\leq\text{TV}(\theta)
\leq H(\theta)\sqrt{2-H(\theta)^2}.
\ee
Thus from $H(\theta)$ we have upper and lower bounds on the optimal validation accuracy. These bounds are plotted in dashed lines in Fig.~\ref{Results1}. We have a guarantee that the optimal performance possible for any network (with any architecture, training time, training method, etc.) lies between the dashed lines. The fact that the network validation accuracy tracks these bounds indicates that the network is near optimal. Moreover, the fact that the upper bound in the signal case (green dashed line) is near to $50\%$ before $\approx15\text{ as}$ indicates that there is no way to process the data to learn where the boundary is before this time. This shows that the lattice approximations used for this example do not violate relativistic causality; no neural network can extract the signal from the data before relativity says it can, simply because the information is not there yet.

It is worth remarking that the network used was not designed with this problem in mind, and yet it performs almost optimally. This suggests that the network is good at extracting and processing all the information contained in the data produced by the measurement protocol. Therefore, it is reasonable to expect a near-optimal behaviour in the other examples as well, especially since we have changed neither the network architecture nor the training procedure. Notice in particular that we did not need to feed the neural network any data about the parameters of the experiment, $\sigma, K, a, L, \omega_\textsc{d}$, and $m$. We did, however, rely on the network being trained on data obtained in the same physical system that it was tested on later. If needed, the network could be trained on data from QFTs with a range of different parameters, so that it also learns to distinguish the dependence on these parameters from the dependence on the feature we are actually interested in. 

\section{A second physical example: Thermometry}\label{Thermometry} 
To showcase the broad applicability of our framework, we consider a very different problem keeping the exact same measurement protocol, the same coupling between probe and field, and the same data-analysis ansatz.

We consider a probe motivated by a superconducting circuit undergoing a long-range interaction with an open transmission line in a thermal state. Such systems do not couple strongly to frequencies above $50\text{ GHz}$~\cite{douglas1964,Forn-Diaz2016,Emma}. Assuming a Gaussian profile we can match this behavior by taking \mbox{$3/\sigma=50\text{ GHz/c}$, i.e., $\sigma=18\text{ mm}$}. These numbers are motivated by~\cite{Emma}. Taking our UV-cutoff in the field at $K=16/\sigma=267\text{ GHz/c}$ gives lattice spacing $a=\pi/K=3.5\text{ mm}$. We couple the circuit to the center of a transmission line of length \mbox{$L=100,\,a=19.6,\,\sigma=353\text{ mm}$}, with Dirichlet boundary conditions. We take the circuit to have an energy gap typical of such systems, $\omega_\textsc{d}=10\text{ GHz}$, and the field to have a mass $m c^2/\hbar = 0.1\text{GHz}$, much smaller than the other energy scales. We again consider  strong-coupling: \mbox{$\lambda_0/\hbar=\omega_\textsc{d}=10\text{ GHz}$}.

\begin{figure}
\includegraphics[width=0.5\textwidth]{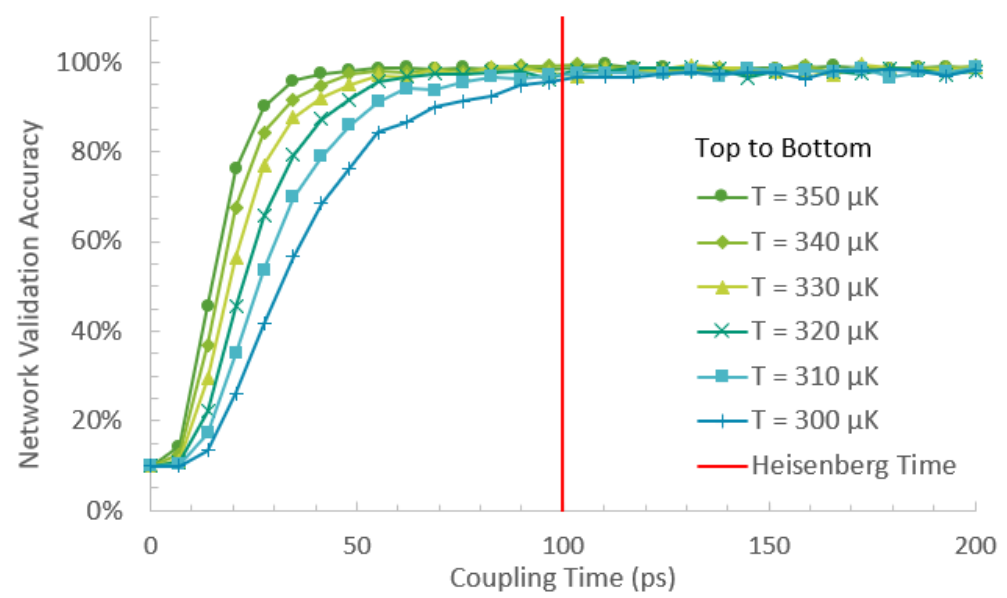}
\caption{A neural network trained to predict (through regression) the temperature of a quantum field from local probe data. The network was trained on labeled data corresponding to field temperatures from a range $T\pm10\%$. The fraction of the validation data which the network labeled correctly to within $\pm1\%$ is plotted as function of the duration of the probe's interaction with the field. A point plotted at time $t$ indicates the network's accuracy given measurements taken at  $N_\textsc{times}=10$ measurement times between $t$ and the previous plot point. The network was trained on $n_\text{train}=7500$ examples from each range. Each example summarizes $N_\textsc{tom}=10^{20}$ measurements of each of the probe's quadratures ($\hat{q}_\textsc{d}$, $\hat{r}_\textsc{d}$ and $\hat{p}_\textsc{d}$) at each measurement time. The vertical red line is the probe's Heisenberg time $\omega_\textsc{d}^{-1}$.}\label{Results2}
\end{figure}

Using these parameters we trained the network to estimate the field's temperature based only on measurements of the local probe. For each base temperature $T$ we generated our labeled data by simulating how the probe would respond to a quantum field of temperature, $y$, selected uniformly from the range $[0.9 T, 1.1T]$, i.e., the range $T\pm10\%$. For each $T$, we trained a neural network through regression to accurately predict the temperature of the field, $y$. To validate the accuracy of the network we determined the fraction of the validation data which the network was able to correctly place the label within $y\pm 0.01 T$ of the correct value. By random chance you would expect the network to guess the correct temperature to within this accuracy $10\%$ of the time. This is what one would expect when the coupling time is zero since the probe has not learned anything about the field. This is confirmed by the $10\%$ validation accuracy shown in Fig.~\ref{Results2} when the coupling time is zero.

As the coupling time increases, the network becomes more accurate. For each base temperature $T$ the network reaches nearly $100\%$ of the validation data labeled correctly (to within $y\pm 0.01 T$). It can do so even before the interaction's thermalization time, which is lower-bounded by the detector's Heisenberg time---the smallest timescale that the detector can resolve--- $1/\omega_\textsc{d}=100\text{ ps}$ (red vertical line). Note that the neural network can determine the temperature very accurately even for very low transmission line temperature (sub-mK). 

Same as in the previous example of boundary sensing, the possibility of measuring temperature without letting the thermometer thermalize is not a new result in itself. It is known that thermometry in times below the thermalization time of the thermometer is possible~\cite{Correa2015,Cavina2018,Boeyens2021,Sekatski2021}. What this example shows is that the field temperature can indeed be reconstructed for times as short as the Heisenberg time of the thermometer using the exact same local measurement protocol and data-analysis ansatz we used for boundary sensing, thus adding temperature to the arguably long list of features of the QFT that can be reconstructed  with our framework using this very innocent choice of data sampling and processing.


\section{A third physical example: Discrimination between Fock and Coherent states}\label{FockExample} 


In this Section we would like to showcase the effectiveness of our framework in an example that assumes neither 1) the Gaussianity of the probe/field states, nor 2) a UV cutoff/bandlimit/lattice discretization. With this in mind, here we apply the proposed measurement framework to the quantum optical problem of distinguishing Fock states (like a single-photon state) from low amplitude coherent states (produced by stimulated emission) when the expectation of the number of photons in the state is the same.

Consider a massless scalar field in a (1+1)-dimensional cavity of length $L$ with Dirichlet boundary conditions. We can consider the mode decomposition of this field, 
\begin{align}
\label{modedecomposition}
\hat{\phi}(t,x)\!=\!
\sqrt{\frac{2\hbar c^2}{L}}
\sum_{\ell=1}^\infty
\frac{\sin(k_\ell x)}{\sqrt{\omega_\ell}}\,
(\hat{q}_\ell\!\cos(\omega_\ell t)
+\hat{p}_\ell\!\sin(\omega_\ell t)),
\end{align}
where the dimensionless quadrature operators $\hat{q}_n$ and $\hat{p}_m$ satisfy canonical commutation relations  $[\hat{q}_n,\hat{p}_m]=\ii\delta_{nm}\openone$ and where $c\,k_\ell=\omega_\ell=\pi\,\ell\,c/L$.

We will take the field state to be the vacuum for all modes except for the lowest frequency one (the $\ell=1$ mode). We will try to determine the initial state of the $\ell=1$ mode by measuring a probe coupled locally to the field in the center of the cavity. We take the $\ell=1$ mode to be in either a) a Fock state $\ket{N}$ with $N$ excitations, or b) a phase-averaged coherent state with $N$ excitations on average. That is, a coherent state $\ket{\alpha}$ for some $\alpha\in\mathbb{C}$ with $\vert\alpha\vert^2=\langle\hat{n}\rangle=N$ but with an unknown phase. In other words, we will consider the two following initial probe-field Wigner functions
\begin{align}
\label{fockAndcohe}
\nonumber
W_1&=W_\text{0}(q_\textsc{d},p_\textsc{d})\!
\times \!W_\text{Fock}(q_1,p_1;N)
\!\times \! \Pi_{\ell=2}^\infty W_\text{0}(q_\ell,p_\ell),\\
W_2&=W_\text{0}(q_\textsc{d},p_\textsc{d})
\!\times \! W_\text{PAC}(q_1,p_1;N)
\!\times \!\Pi_{\ell=2}^\infty  W_\text{0}(q_\ell,p_\ell)
\end{align}
where $q_\textsc{d}$ and $p_\textsc{d}$ are the probe variables and where $W_\text{0}(q,p)= e^{-q^2-p^2}/\pi$. Note that neither of these states are Gaussian. The Wigner function of a Fock state is~\cite{Cahill1969},
\begin{align}
W_\text{Fock}(q,p;N)&=\frac{(-1)^{N}}{\pi} L_N(2(q^2+p^2)) \ e^{-(q^2+p^2)}
\end{align}
where $L_N(x)$ is the $N$-th Laguerre polynomial. For the unknown phase coherent state, the fact that we do not know (therefore average over) the phase makes this a non-Gaussian state. The Wigner function of a phase-averaged coherent state (PAC) is~\cite{Allevi2013} 
\begin{align}
W_\text{PAC}(q,p;N)&\!=\!
\!\int_0^{2\pi}\!\!\frac{\d\theta}{2\pi^2}
 \ \!e^{-(q-\sqrt{2N}\cos(\theta))^2
-(p-\sqrt{2N}\sin(\theta))^2}.
\end{align}
Moreover, we note that these two states have exactly the same first and second moments:
\begin{align}
&\langle\hat{q}\rangle_\text{Fock;N}
=\langle\hat{p}\rangle_\text{Fock;N}
=0,
\quad
\langle\hat{q}\,\hat{p}\rangle_\text{Fock;N}=0,
\\
\nonumber &\langle\hat{q}^2\rangle_\text{Fock;N}
=\langle\hat{p}^2\rangle_\text{Fock;N}
=\langle\hat{n}+1/2\rangle_\text{Fock;N}
=N+1/2,
\end{align}
and
\begin{align}
&\langle\hat{q}\rangle_\text{PAC;N}
=\langle\hat{p}\rangle_\text{PAC;N}
=0,
\quad
\langle\hat{q}\,\hat{p}\rangle_\text{PAC;N}=0,
\\
\nonumber &\langle\hat{q}^2\rangle_\text{PAC;N}
=\langle\hat{p}^2\rangle_\text{PAC;N}
=\langle\hat{n}+1/2\rangle_\text{PAC;N}
=N+1/2.
\end{align}
Thus no analysis of these two field states in terms of their first and second moments can differentiate them; these field states can only be distinguished by methods which are sensitive to their third and higher order moments.

\begin{figure*}
\includegraphics[width=0.54\textwidth,height=0.35\textwidth]{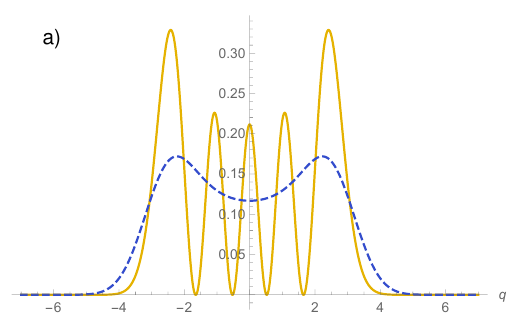}
\includegraphics[scale=1.1]{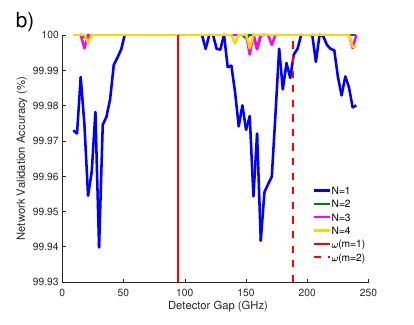}
\caption{(Color online.) a) The marginal distributions for the $N=4$ Fock state (solid) and for the phase averaged coherent state with $\vert\alpha\vert^2=4$ (dashed). The distributions are not Gaussian and they have the same mean and variance, making them impossible to distinguish with simple statistical analysis of  first and second moments. 
b) The validation accuracy of a neural network trained to distinguish two field states from the measurements of a local detector coupled to the field. In particular the network differentiates vacuum cavity states with the following two modifications: 1) the lowest field mode is in an $N$-particle Fock state or 2) the lowest field mode is in a coherent state with expectation $\langle\hat{n}_1\rangle=N$ particles and unknown phase.
}\label{Marginals}
\end{figure*}

\begin{figure*}
\includegraphics[scale=1.15]{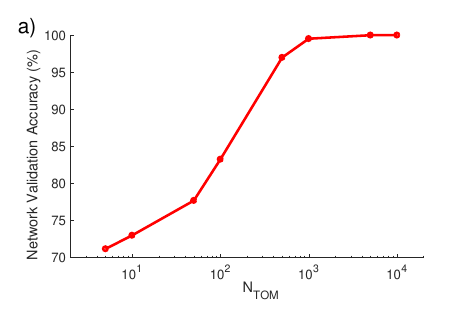}
\includegraphics[scale=1.15]{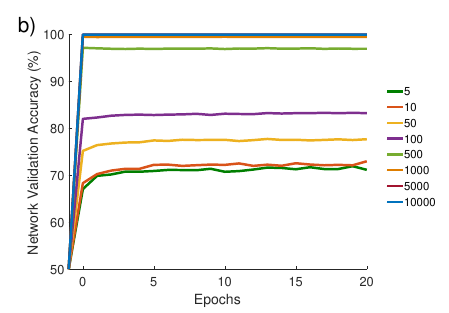}
\caption{(Color online.) a) Validation accuracies attained for different tomographic sizes $N_\textsc{tom}$, for expected number of particles $N=1$, and detector frequency $\omega_\textsc{d}=\omega_1$. b) Evolution of network validation accuracies with the number of training iterations, plotted for different tomographic sizes. As before, the expected number of particles is $N=1$, and the frequency of the detector is that of the first mode of the lattice, $\omega_\textsc{d}=\omega_1$. 
}\label{MoreintoFockvsAver}
\end{figure*}

It is a non-trivial task to determine which of these states the field is in. Suppose that (forgoing the localized probe system temporarily) we are somehow able to measure one of the quadrature operators of the lowest mode (e.g., $\hat{q}_1$) directly. The outcome of this measurement would be selected from the marginal distributions of $W_\text{Fock}(q,p;N)$ and $W_\text{PAC}(q,p;N)$. These are shown in Fig. \ref{Marginals}a for the case of $N=4$. The total variation distance between these marginals is $\text{TV}\approx 0.29$, such that best odds one can hope for given a single measurement outcome are $(1 + \text{TV})/2\approx 64.5\%$. 

In actuality we will attempt to distinguish these field states from a harmonic oscillator probe's response to the field, as explained in Sec.~\ref{GeneralFramework}. This probe will pick up information from all of the field modes, much of which is irrelevant to the task at hand. Our machine learning algorithm will need to learn to distinguish the irrelevant noise 
from the (already weak) signal from the $\ell=1$ mode.

 The only relevant difference with the measurement procedure in Secs.~\ref{BoundarySensing} and~\ref{Thermometry} is the switching function. While the same results would be obtained with the same coupling protocol as in the previous section, for ease of analytical treatment in this manuscript we consider a switching function $\chi(t)=\delta(t)+\delta(t-t_\textsc{m})$ where $t_\textsc{m}$ is a time just before we measure the probe. Specifically, the probe undergoes a strong sudden interaction with the field $t=0$. Then both the probe and field evolve freely for a time $t_\textsc{m}$. The probe undergoes another strong sudden interaction with the field at $t=t_\textsc{m}$. Finally, we measure one of the probe operators ($\hat{q}_\textsc{d}$, $\hat{p}_\textsc{d}$ or $\hat{r}_\textsc{d}$). This measurement procedure is repeated $N_\textsc{tom}$ times for each probe operator and at each of the $N_\textsc{times}$ measurement time $t_\textsc{m}$. It is important to note that since the field state is not Gaussian, these probe measurement values will not be distributed normally. The distributions they are drawn from are ultimately derived from the ones in Fig. \ref{Marginals}a and are much noisier due to vacuum noise from the other field modes that also couple to the probe (we do not carry out any single mode or rotating-wave approximations).

As in the boundary sensing and thermometry examples, we record the sample means and sample variances of these $N_\textsc{tom}$ measurements in our compressed data. However, as discussed above, we will need more than just first and second moments to handle this problem. Thus we additionally include the central fourth moments of the distribution of sampled data, as explained in Sec.~\ref{method2}. In Appendix \ref{appendix4Moments} and \ref{appendifxFockCoherent} we calculate non-perturbatively the second, fourth and eighth moments of the probe's quadrature operators for the two field states that are relevant for the example in Sec.~\ref{FockExample}. Recall that we do not require any more measurements, we just need the network to be able to process higher moments from the same sample of measurements used previously (i.e., less compression). We then train our neural network on many examples of this compressed data until it can accurately classify whether any given data came from an interaction with a Fock state or a coherent state. We use exactly the same neural network architecture, loss function and optimization method as in the two previous examples. 

We consider an optical cavity of length $L=1 \text{ cm}$ and a probe with a Gaussian smearing function of width $\sigma=0.1 \text{ mm}$ placed at the center of the cavity \mbox{$x = L/2 = 0.5 \text{ cm}$}. As discussed above, in each run of the experiment the probe strongly interacts with the cavity at two times: first at $t=T_{\textsc{min}}=0$, then at $t = T_{\textsc{min}} + n \,\Delta t$ for $n = 1, 2, \ldots, N_\textsc{times}$ with $\Delta t = 6.67\text{ ps}$. As $N_{\textsc{times}}=10$, we have that $T_{\textsc{max}}=66.7\text{ ps}$.

Fig.~\ref{Marginals}b shows the validation accuracy of the neural network given different values for the probe frequency, $\omega_{\textsc{d}}$, and for the same tomography, $N_{\textsc{tom}}=5000$. This shows that the neural network can successfully distinguish between a Fock and a PAC state given the sample fourth moments. The neural network can distinguish the two states with almost 100\% accuracy in a wide range of detector gaps. This plot also provides some physical insight: at resonance, i.e. when the probe frequency is the same as the frequency of the mode of interest (in Fig.~\ref{Marginals}b, a solid vertical line corresponds to \mbox{$\omega_\textsc{d} = \omega_1 = \pi/L = 94.2\text{ GHz}$}), we obtain an improvement in the accuracy of the neural network. Furthermore, there is a second peak at double the frequency of the first mode (in Fig. \ref{Marginals}b, the dashed vertical line corresponds to \mbox{$\omega_\textsc{d} = 2 \,\omega_1=188.4\text{ GHz}$}). This is expected since when the mode frequency is an integer multiple of the detector energy gap, resonance occurs and the detector is more sensitive to getting excited by capturing field excitations. We also show in Fig.~\ref{MoreintoFockvsAver} how the accuracy a) increases rapidly with the number of measurements, and b) how it evolves during training. It is worth noticing that as the number of measurements increases, the neural network becomes increasingly fast at reaching a stable validation accuracy. Fig.~\ref{MoreintoFockvsAver}a also shows that the validation accuracy almost saturates with a moderate number of measurements ($N_\textsc{tom}\sim 10^4$), which supports the experimental viability of the proposal. In previous examples, we considered particularly big tomographic sizes ($N_\textsc{tom}=10^{22}$ in Sec.~\ref{BoundarySensing}, and $N_\textsc{tom}=10^{20}$ in Sec.~\ref{Thermometry}) to ensure that the validation accuracy that we analyzed was the sole result of the network's ability to unscramble the information from the local measurements, and that no role was played by the possible inaccuracies of the data it was fed with---even though these potential inaccuracies and the resilience of the protocol against them do play an important role in experiments. Moreover, it should be taken into account that the number of measurements $N_\textsc{tom}$ does not necessarily translate into \textit{sequential} repetitions of the measurement protocol. For instance, collective measurements made on ensembles of particles (e.g., an atomic gas) can be translated into averages of individual observables calculated over the number of particles of the ensemble, which typically is already of the order of the Avogadro number ($N_\textsc{a} \sim 10^{24}$). In this kind of setup, tomographic sizes such as those considered in the two previous examples are clearly within the reach of experiments. 

The success of the measurement framework in this last example shows that its applicability is not restricted to simple Gaussian systems. Indeed, we emphasized that here we have recovered a feature of the field without using the Gaussianity of the probe/field states or a UV cutoff/bandlimit/lattice discretization.

\section{Conclusions}

Local measurements of a quantum field can reveal information about its global features. We have shown that we can use machine learning techniques to unscramble the information about QFTs acquired by localized probes with a one-size-fits-all method, thus avoiding the necessity of designing a specific measurement protocol and data-analysis function for each feature we might be interested in. More concretely, we have demonstrated how to read out non-local features of a QFT from the outcomes of a fixed measurement protocol with local experiments, processed through a neural network with a generic architecture and training procedure.

As particular examples to showcase the power of the proposed machine learning framework we have examined three case studies: i) how a local probe can see a wall far away from it, in the vacuum and without actively sending signals to bounce off it, ii)  how a local probe that is not given enough time to thermalize can still accurately determine the temperature of a quantum field, and iii) how detectors can accurately distinguish between Fock states and coherent states even when the first and second statistical moments of their observables match. To do so, in all cases we used the same simple measurement protocol, which was not adapted to the particular toy problems considered in this paper. Yet we were able to distinguish with high levels of accuracy the relevant features of the field we were after in each case. This is evidence of the potential of these methods to accommodate experimental needs. Namely, the use of machine learning techniques in the context of quantum field theory takes the complexity burden out of the design of experimental protocols and puts it on the data processing, which neural networks can deal with efficiently.


The techniques we present in this paper are general and of wide applicability. This paves the way to the use of machine learning techniques in more complicated scenarios such as distinguishing gravitational backgrounds~\cite{keithng,Smith_2014}, global state tomography~\cite{Torlai2018} with local probes, acknowledging entanglement in analogue Hawking radiation~\cite{Steinhauer2016}, and maybe even new experimental proposals seeking direct evidence of yet untested QFT phenomena such as the Unruh effect~\cite{unruh1976}. In each of these scenarios the response of local probes like the ones in this paper are often used to study features of the QFT, so the techniques proposed in this manuscript are directly applicable. Finally, the methods developed here are directly translatable to their use in many-body quantum physics, where they can be used to address the problem of measuring many-body observables with local probes in, e.g., quantum phase transitions~\cite{Xu_2019}.

\begin{acknowledgments}
The authors would like to thank Luis J. Garay for enlightening discussions. DG acknowledges support by NSERC through a Vanier Scholarship. JPG acknowledges support by a Mike and Ophelia Lazaridis Fellowship. JPG also acknowledges the support of a fellowship from ``La Caixa'' Foundation (ID 100010434, code LCF/BQ/AA20/11820043). EMM acknowledges support through the Discovery Grant Program of the Natural Sciences and Engineering Research Council of Canada (NSERC). EMM also acknowledges support of his Ontario Early Researcher award. Research at Perimeter Institute is supported in part by the Government of Canada through the Department of Innovation, Science and Industry Canada and by the Province of Ontario through the Ministry of Colleges and Universities. This work was made possible by the facilities of the Shared Hierarchical 
Academic Research Computing Network (SHARCNET:www.sharcnet.ca) and Compute/Calcul Canada. 
\end{acknowledgments}


\appendix
\onecolumngrid
\section{Preprocessing, Neural Network Architecture and Training Details}\label{appendixA}

As we described in the main text, our measurement procedure and compression produces labeled data consisting of data $D_\textsc{c}\in\mathbb{R}^d$ where $d=9\, N_\textsc{times}$ and an associated label $y$. To begin training we collect $n=N_\textsc{samples}$ instances of this labeled data into a $d\times N_\textsc{samples}$ design matrix $\bm{X}=(D_1,\dots,D_n)^\intercal$ and a vector of labels $\bm{y}=(y_1,\dots,y_n)^\intercal$. We then portion off $75\%$ of this data ($N_\textsc{train}=0.75\,N_\textsc{samples}$) to be used for training the neural network, $\bm{X}_\textsc{train}$ and $\bm{y}_\textsc{train}$, leaving the other $25\%$ ($N_\textsc{valid}=0.25\,N_\textsc{samples}$) as validation data, $\bm{X}_\textsc{valid}$ and $\bm{y}_\textsc{valid}$, which we will ultimately use to test the accuracy of the trained network. Note that the network will not be exposed to any of the validation data during training.

We begin processing our data by subtracting off the mean of the training data, $\bm{X}\to\bm{X}-\bm{X}^\text{avg}_\textsc{train}$, where \mbox{$\bm{X}^\text{avg}_\textsc{train}
=\sum_{k=1}^{N_\textsc{train}}
D_k/ N_\textsc{train}$}. Next we do principle component analysis (PCA), which finds a representation of our data without linear correlations. To do this we compute the covariance matrix of our training data and perform a singular value decomposition on it,
\be
\frac{1}{N_\textsc{train}-1}\bm{X}_\textsc{train}^\intercal \bm{X}_\textsc{train}
=V^\intercal \Lambda V
=\sum_{j=1}^{d}
\lambda_j \, \bm{v}_j\bm{v}_j^\intercal
\ee
where $V=(\bm{v}_1,\dots,\bm{v}_{d})^\intercal$ is the matrix of singular vectors, $\bm{v}_j$, and $\Lambda = \text{diag}(\lambda_0,\dots\lambda_{d})$ is the matrix of singular values, $\lambda_j\in\mathbb{R}_+$. The singular vectors are the directions in which our data varies independently, and the singular values indicate ``how much'' variance is in each direction. Using this decomposition we can rewrite our data in this singular basis by taking $\bm{X}\to V\bm{X}$. After this transformation, the training data has a diagonal covariance matrix, namely $\Lambda$. Finally we can whiten the data by taking  $\bm{X}\to \Lambda^{-1/2}\bm{X}$. The covariance matrix of the training data is now the identity matrix. We do this in order to force the neural network to take into account all components of the data, since they come from sources that are not supposed to be directly comparable in magnitude. Note that we have not used PCA to compress our data; that is, we have not cut any small singular values out of $\Lambda$ as is commonly done.

The data is now ready to begin training the neural network. As discussed in the main text, neural networks work by alternatingly applying tuneable linear-affine transformations (controlled by weights and biases) and fixed non-linear transformations (the activation function) to their inputs. See Fig. \ref{NN} for a schematic of a neural network that can be used to classify the topology of a QFT based on local probe measurement data.

We will now use the architecture in Fig. \ref{NN} as a basic illustrative example. In this example the network accepts a 5-dimensional input, $\bm{x}^{(0)}$, into the left-most layer of the network (note that in the examples discussed in the main text the input dimension is much larger). In passing this data to the next layer of the network, a linear-affine transformation is applied to $\bm{x}^{(0)}$, as $\bm{x}^{(1)}=A^{(1)}\bm{x}^{(0)}+\bm{b}^{(1)}$. The weight matrix $A^{(1)}$ here has dimensions $7\times5$ and the bias vector has a dimension of $7$ such that $\bm{x}^{(1)}$ is 7-dimensional. The $7\times5+7=42$ values which determine this linear-affine transformation are left as free parameters to be optimized during training. Next, a fixed non-linear function, $\mathcal{G}^{(1)}$ is applied element-wise to each entry of $\bm{x}^{(1)}$ yielding \mbox{$\bm{z}^{(1)}=\mathcal{G}^{(1)}(\bm{x}^{(1)})$}. For instance $\mathcal{G}^{(1)}$ may be the hyperbolic tangent function or a rectified linear unit.

This process is then repeated at each layer. First a linear affine transformation is applied to $\bm{z}^{(1)}$ as \mbox{$\bm{x}^{(2)}=A^{(2)}\bm{z}^{(1)}+\bm{b}^{(2)}$} where $A^{(2)}$ has dimensions $5\times7$ and $\bm{b}^{(2)}$ has dimension $5$. Then a fixed non-linear function, $\mathcal{G}^{(2)}$ is applied element-wise to $\bm{x}^{(2)}$ yielding \mbox{$\bm{z}^{(2)}=\mathcal{G}^{(2)}(\bm{x}^{(2)})$}. In the final layer we have \mbox{$\bm{x}^{(3)}=A^{(3)}\bm{z}^{(2)}+\bm{b}^{(3)}$} where $W^{(3)}$ has dimensions $2\times5$ and $\bm{b}^{(3)}$ has dimension $2$ and \mbox{$\bm{z}^{(3)}=\mathcal{G}^{(3)}(\bm{x}^{(3)})$} for some non-linear function, $\mathcal{G}^{(3)}$. In total this network computes the function $f(\bm{x}^{(1)};A,b)=\bm{z}^{(3)}$ where $A$ and $b$ refer to this network's $94$ free parameters collectively.

There are two different problem types we need to design a network for, classification and regression. In classification, our network is tasked with deciding which of several classes (given by a discrete label $y$) our data belongs to. In this scenario we take the number of neurons in the final layer to be equal to the number of classes, and the final activation function to be a softmax. This ensures that the network's final output is a probability distribution that can be interpreted as the probability that the initial data belongs to each class. In regression, our network is tasked with assigning the data a continuous label $y$. In this case we take the final layer to have a single neuron.

In the examples discussed in the paper we considered a network consisting of 90 neurons on the input layer, 30 in the intermediate (hidden) layer and either two or one neurons in the final layer depending on which example we are doing. In the boundary sensing and state discrimination examples we have two neurons in the final layer. In the thermometry cases we have only one neuron in the final layer. All of the non-linear activation functions were taken to be leaky rectified linear units~\cite{Goodfellow}.

The network's weights and biases are tuned to minimize error of the network's predictions over the training set. To quantify this error we define the following cost functions,
\begin{align}
\text{Classification:}\qquad &C(A,b)=\frac{-1}{N_\textsc{train}}\sum_{k=1}^{N_\textsc{train}} \ \tilde{y}_k \cdot \text{log}(f(\bm{x}_k;A,b))\\
\text{Regression:}\qquad &C(A,b)=\frac{1}{N_\textsc{train}}\sum_{k=1}^{N_\textsc{train}} \ (f(\bm{x}_k;A,b)-y_k)^2
\end{align}
where $\tilde{y}_k$ is the one-hot encoding of the $k^{th}$ data point's label. For the classification scenario, our cost function is the relative entropy between the network's probability assignment and the expected result. For the regression case, the cost function is the mean square error. To help reduce overfitting (that is, an excessively close alignment of the network's model to the training data that might end up worsening its performance with real---or validation---data) we add an $L_2$ regularizer to this cost function, $\lambda_2\vert\vert A\vert\vert_2^{2}$, for some chosen $\lambda_2$. This reduces the complexity of the model by penalizing the network for using large weights. Additionally, when training the network, we randomly ``drop'' some fraction of the neurons. This forces the network to be more robust. The sum of the cost function and the regularizer are then minimized by stochastic gradient descent~\cite{Goodfellow}.

\section{Free scalar QFT on the lattice}\label{appendixB}
As we discussed in the main text, we can motivate a UV-cutoff for the field-probe system through the length scale of the probe's smearing function. To see this, let us expand the field-probe interaction Hamiltonian
\begin{align}
\hat{\mathcal{H}}_\textsc{int} = \lambda \, \chi(t) \int_{-\infty}^{\infty} \d x \, F(x) \, \hat{q}_\textsc{d}(t) \otimes \hat{\phi}(t,x)
\end{align}
in terms of plane-wave modes as
\begin{align}
\hat{\mathcal{H}}_\textsc{int} 
&= \lambda \, \chi(t)\, \hat{q}_\textsc{d}(t)  \int_{\mathbb{R}^2} \frac{\d x \, \d k}{2\sqrt{\pi\omega_k}} \, F(x) \left(e^{-\ii \omega_k t} e^{\ii k x} \hat{a}_k +\text{H.c.}\right)
= \lambda \, \chi(t)\, \hat{q}_\textsc{d}(t) \int_{-\infty}^\infty \frac{\d k}{2\sqrt{\pi\omega_k}} \,\left(\tilde{F}(k)\,e^{-\ii \omega_k t}\hat{a}_k +\text{H.c.}\right),
\end{align}
where $\tilde{F}(k)=\mathcal{F}_k[F(x)]$ is the Fourier transform of $F(x)$. Note that $\tilde{F}(k)$ determines how strongly the probe couples to each of the field modes. If the smearing function decays fast enough outside of a region of size $\sim\sigma$ (e.g., $F(x)$ is a Gaussian with standard deviation $\sigma$) then $\tilde{F}(k)$ would have an approximate width $\sim1/\sigma$. That is, the probe would not couple strongly to modes with wavenumber $\vert k\vert\gg \sigma^{-1}$. Thus by considering a probe with an effectively finite spatial extent  we are automatically considering a soft-UV-cutoff in the interaction of field and probe.

If $\tilde{F}(k)$ decays sufficiently fast, we can neglect the coupling to the modes above some large UV threshold, say $\vert k\vert > K$ (e.g. for a Gaussian profile we can take $K=16/\sigma$). This yields an effective coupling of \mbox{$\tilde{F}^\textsc{uv}(k)
\coloneqq \Pi_K(k)\tilde{F}(k)$} where $\Pi_K(k)$ is the rectangle function over $k\in[-K,K]$. By the Nyquist-Shannon sampling theorem we can then reconstruct our UV-cutoff smearing function, \mbox{$F^\textsc{uv}(x)\coloneqq\mathcal{F}_{x}^{-1}[\tilde{F}^\textsc{uv}(k)]$}, as \mbox{$F^\textsc{uv}(x)= \sum_n F^\textsc{uv}(x_n) \,S_n\!\left(x/a\right)$}, where $a=\pi/K$ is the spacing of the discrete positions, $x_n = n\,a$, and where $S_n(r)\coloneqq\sin(\pi (r-n))/\pi (r-n)$ is a displaced normalized sinc function.

Note that in general \mbox{$F^\textsc{uv}(x_n) \neq F(x_n)$}. This means that in order to recover the UV-cutoff smearing function we cannot sample the original smearing function, but its bandlimited version instead. However, precisely because we are assuming that $\tilde{F}(k)$ is effectively bandlimited, we can approximate $F^\textsc{uv}(x_n) \simeq F(x_n)$. Indeed, for the particular case of a Gaussian with standard deviation $\sigma$ and $K=16/\sigma$, it can be straightforwardly shown that \mbox{$|F(x)-F^\textsc{uv}(x)| \lesssim 10^{-59}$} for every real $x$. Now, note that $S_n(r)$ decays only polynomially for large $r$. Thus in general our UV-cutoff smearing function will have polynomial tails, as all bandlimited functions do. This might seem to be in contradiction with our previous approximation, since $F^\textsc{uv}(x_n)$ will decay polinomially, while $F(x_n)$ does so exponentially. However, the bound above shows precisely that these differences in the rythm of decay are only relatively significant in the regions in which the order of magnitude of both $F(x)$ and $F^\textsc{uv}(x)$ is already negligible. We can therefore faithfully approximate the UV-cutoff smearing function by sampling $F(x)$ instead of $F^\textsc{uv}(x)$, i.e., we redefine \mbox{$F^\textsc{uv}(x) = \sum_n F(x_n) \,S_n\!\left(x/a\right)$}. This function is still bandlimited and so still has polynomial tails, however the coefficients $F(x_n)$---which, as we will soon see, tell us how the probe couples to the lattice sites---are sampled directly from the original smearing function.

Since $F^\textsc{uv}$ as defined above is bandlimited, we can define the UV-cutoff interaction Hamiltonian as
\begin{align}
\hat{\mathcal{H}}_\textsc{int}^\textsc{uv} 
\coloneqq \lambda \, \chi(t) \int_{-\infty}^{\infty} \d x \, F^\textsc{uv}(x) \, \hat{q}_\textsc{d}(t) \otimes \hat{\phi}(t,x)
=\lambda \, \chi(t) \int_{-\infty}^{\infty} \d x \, F^\textsc{uv}(x) \, \hat{q}_\textsc{d}(t) \otimes \hat{\phi}^\textsc{uv}(t,x)
\end{align}
where we note that the UV-cutoff smearing function effectively induces a UV-cutoff of the field operator,
\begin{equation}
\hat{\phi}^\textsc{uv}(t,x)\coloneqq\mathcal{F}^{-1}_x\left[\Pi_K(k)\mathcal{F}_k[\phi(t,x)]\right]= \int_{-K}^{K} \frac{\d k}{2\sqrt{\pi\omega_k}} \,\left(\hat{a}_k e^{-\ii \omega_k t} e^{\ii k x} +\text{H.c.}\right) 
\end{equation} 
Next, we note that since $\hat{\phi}^\textsc{uv}(t,x)$ is bandlimited we can express it as a sum of $\text{sinc}$ functions as, \mbox{$\hat{\phi}^\textsc{uv}(t,x) = \sum_n \hat{\phi}^\textsc{uv}(t,x_n) \, S_n\!\left(x/a\right)$}. Recomputing the UV-cutoff interaction Hamiltonian using these sinc representations we find
\begin{align}\label{HamilIntUV}
\hat{\mathcal{H}}^\textsc{uv}_\textsc{int} = \lambda \, \chi(t) \sum_n \, a \, F(x_n) \, \hat{q}_\textsc{d}(t) \otimes \hat{\phi}^\textsc{uv}(t,x_n)
\end{align}
where we have used the orthonormality of the collection $\{S_m(r)\}$ in the $L^2$ norm. Thus, by taking a hard UV-cutoff on the probe's smearing function we automatically find that the probe effectively only couples to the field at the discrete positions $x_n = n \,a$.

Notice that so far we are not implying that the field itself has a UV-cutoff or that the space it lives on is discretized. We have only discussed an approximation of the probe coupling. We could study the field theory as is without an explicit UV-cutoff, but for our purposes it is convenient to consider that the field is also bandlimited. We apply this UV-cutoff to the field by removing the field modes with $k>\vert K\vert$, yielding
\begin{align}
\hat{\phi}^\textsc{uv}(t,x)&\coloneqq\mathcal{F}_x^{-1}\left[\Pi_K(k)\mathcal{F}_k[\hat\phi(t,x)]\right],\\
\nonumber
\hat{\pi}^\textsc{uv}(t,x)&\coloneqq\mathcal{F}_x^{-1}\left[\Pi_K(k)\mathcal{F}_k[\hat\pi(t,x)]\right],\\
\nonumber
\partial_x\hat{\phi}^\textsc{uv}(t,x)&\coloneqq\mathcal{F}_x^{-1}\left[\Pi_K(k)\mathcal{F}_k[\partial_x\hat\phi(t,x)]\right].
\end{align}
Note that since these operators are now bandlimited we can express them as
\begin{align}\label{bandlimited QFT}
\hat{\phi}^\textsc{uv}(t,x) 
&=\sum_n \hat{\phi}^\textsc{uv}(t,x_n) \, S_n\!\left(x/a\right),\\
\nonumber
\hat{\pi}^\textsc{uv}(t,x) 
&=\sum_n \hat{\pi}^\textsc{uv}(t,x_n) \, S_n\!\left(x/a\right),\\
\nonumber
\partial_x\hat{\phi}^\textsc{uv}(t,x)
&=\sum_n \partial_x\hat{\phi}^\textsc{uv}(t,x_n) \, S_n\!\left(x/a\right).
\end{align}
The UV-cutoff field Hamiltonian is then the free field Hamiltonian for this bandlimited QFT, namely
\begin{align}\label{HamilUV}
\hat{\mathcal{H}}_\phi^\textsc{uv}
&\coloneqq \dfrac{1}{2}\int_{-\infty}^{\infty} \d x \, \Big(c^2 \hat \pi^\textsc{uv}(t,x)^2 + (\partial_x\hat\phi^\textsc{uv}(t,x))^2+ \dfrac{m^2 c^2}{\hbar^2} \hat \phi^\textsc{uv}(t,x)^2 \Big) \\
&=\dfrac{a}{2}\sum_n    \Big( c^2\hat{\pi}^\textsc{uv}(t,x_n)^2 
+(\partial_x\hat{\phi}^\textsc{uv}(t,x_n))^2
+\dfrac{m^2c^2}{\hbar^2} \hat{\phi}^\textsc{uv}(t,x_n)^2 \Big) \;, \label{UV Hamiltonian non-lattice}
\end{align}
where we have again used the operator's sinc representations and the $L^2$ orthonormality of $\{S_m(r)\}$ to express the integral as a sum. This way, we have \textit{completely} reduced the dynamics to the field amplitudes and momenta at points $(t,x_n)$. One may think that the fact that the Hamiltonian $\hat{\mathcal{H}}_\phi^\textsc{uv}$ has terms with the field derivatives $\partial_x \hat\phi^\textsc{uv}(t,x_n)$ shows that we are still dealing with a continuum space, not a lattice. Surprisingly this is not the case, these continuous derivative terms are understandable in terms of the discrete lattice with no approximation. 

To see this note that as discussed above, bandlimited function can be perfectly represented on a lattice. The derivative of a bandlimited function is itself a bandlimited function. Thus, for bandlimited functions, derivatives are perfectly understandable on a lattice. This is facilitated by the following remarkable derivative approximation (which is exact for bandlimited functions):
\begin{align}\label{ExactDerivative}
\partial_x f(x)
&= 2\sum_{m=1}^\infty (-1)^{m+1} \frac{f(x+m\,a)-f(x-m\,a)}{2\,m\,a}.
\end{align}
Namely, when $f$ is bandlimited with bandwidth of $K$ and $a\leq\pi/K$ then this formula for the derivative is exact. Moreover, if the Fourier transform of $f$ is mostly supported in $[-K,K]$ with thin tails (e.g, Gaussian tails) outside this region, then this is a very good derivative approximation.

We can apply this logic to field operator as well. By Eq.~\eqref{bandlimited QFT} we have,
\begin{equation}
\partial_x \hat\phi (t,x)=\sum_n \partial_x \hat\phi(t,x_n) \,S_n(x/a)=\sum_n \hat\phi(t,x_n) \, \partial_x S_n(x/a) \;.
\end{equation}
The derivative of the sinc profile $\partial_x S_n(x/a)$ is bandlimited and so can be written as a sum of sinc profiles. Carrying this out we find for $k\in\mathbb{Z}$, we get
\begin{equation}\label{derivative exact}
\partial_x\hat\phi(t,x_k)=\frac{1}{a}\sum_{n\neq k} \frac{(-1)^{k-n}}{k-n} \hat\phi(t,x_n) \;,
\end{equation}
where we have used the $L^2$ orthonormality of $\{S_m(r)\}$ again. Thus, we see that the derivative $\partial_x \hat\phi(t,x_k)$ has an expression in terms of the field operators at $(t,x_n)$. 

If we use Eq.~\eqref{derivative exact} in Eq.~\eqref{UV Hamiltonian non-lattice}, we get an expression for the UV-cutoff Hamiltonian that is fully defined within the 1D lattice $\{x_n\}$. Our bandlimited QFT is perfectly lattice-representable. Indeed as discussed in \cite{DiscreteGenCovPart2}, despite what you may have heard, there are perfectly Lorentzian lattice theories. 

Note, however, that the derivative understood in terms of the lattice sites is in a sense extremely non-local: it involves \textit{all} $n\neq k$. How is it that a perfectly local operation in the continuum (i.e., differentiation) is here being exactly represented by a non-local operation on the lattice sites? This issue is discussed at length in \cite{DiscreteGenCovPart1}, but the ultimate resolution is as follows: the lattice site themselves ought to be thought of as non-local objects, each associated with overlapping sinc-profiles. Thus, our perfectly local differentiation in the continuum is carried out in terms of the lattice via a non-local combination of non-local objects.

If, however, we want to think of the lattice sites themselves as being local objects undergoing nearest-neighbour interactions, then the dynamics must be modified further. To achieve this instead of the exact formula Eq.~\eqref{derivative exact} we instead take the approximation, $\partial_x\hat{\phi}(t,x_n)
\approx [{\hat{\phi}(t,x_{n+1}) -\hat{\phi}(t,x_n)}]/a$, yielding
\begin{align}
\hat{\mathcal{H}}_\phi^\textsc{uv}
&\approx\dfrac{a}{2}\sum_n c^2\hat{\pi}^\textsc{uv}(t,x_n)^2 
+\left(\frac{\hat{\phi}^\textsc{uv}(t,x_{n+1}) - \hat{\phi}^\textsc{uv}(t,x_n)}{a}\right)^2
+\dfrac{m^2c^2}{\hbar^2} \hat{\phi}^\textsc{uv}(t,x_n)^2.
\end{align}
We note that these satisfy the commutation relations, \mbox{$[\hat{\phi}^\textsc{uv}(t,x_n),\hat{\pi}^\textsc{uv}(t,x_m)]=\ii\hbar( \delta_{nm}/a)\openone$}. Finally, rewriting this Hamiltonian in terms of the dimensionless operators, \mbox{$\hat{q}_n = \sqrt{a m/\hbar^2} \, \hat{\phi}^\textsc{uv}(t,x_n)$} and
\mbox{$\hat{p}_n = \sqrt{a/m} \, \hat{\pi}^\textsc{uv}(t,x_n)$} which satisfy the commutation relations, \mbox{$[\hat{q}_i,\hat{p}_j]=\ii \delta_{ij}\openone$},
yields the bandlimited field Hamiltonian of Eqs.~\eqref{HPhiUV} and~\eqref{HIntUV}.

\begin{figure*}
\includegraphics[width=0.45\textwidth]{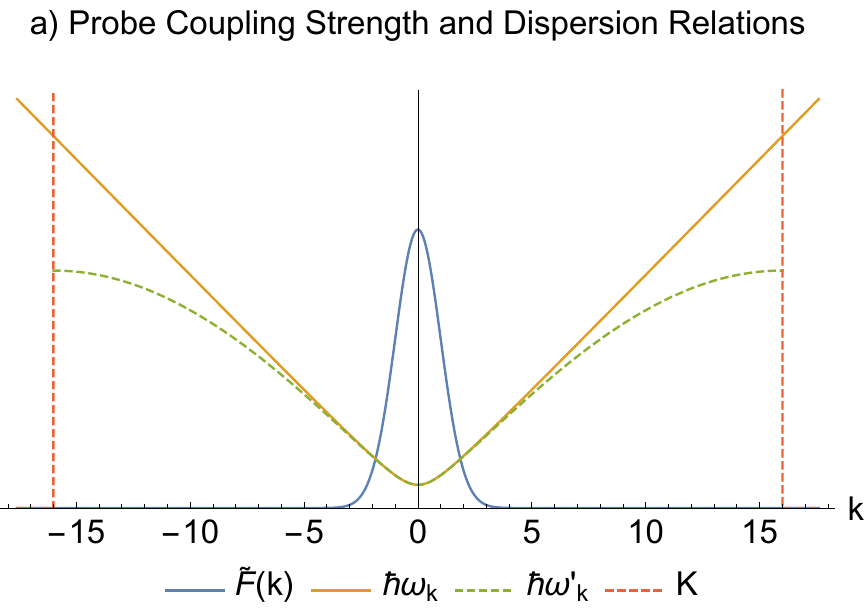}
\includegraphics[width=0.48\textwidth,height=0.32\textwidth]{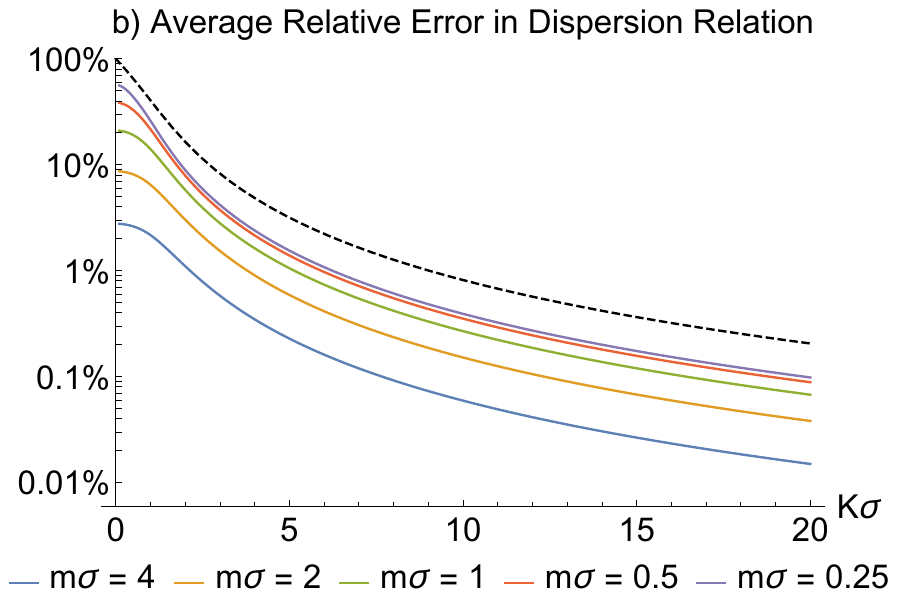}
\caption{In subfigure a) we show the probe's coupling strength to the field modes, $\tilde{F}(k)$, (blue Gaussian) as a function of the mode's wavenumber, $k$. Note that the probes width is taken to be $\sigma=1$. The field's dispersion relation $\hbar\omega_k$ is also plotted (yellow hyperbola). Note that the field's mass is taken to be $m=1$. Taking a UV-cutoff at $K=16$ (vertical red dashed line) yields a modified dispersion relation, $\hbar\omega'_k$, (green dashed) at high frequencies. In subfigure b) we plot the average relative error in $\hbar\omega_k$ as a function the cutoff $K$ and the field mass $m$. This error decreases polynomially as $K\sigma$ increases. The error also decreases as the mass of the field increases. The black dashed line is a mass independent upper-bound on this error. In both subfigures we have taken $\hbar=c=1$.}\label{Dispersion}
\end{figure*}

The implementation of this bandlimitation has introduced essentially three changes to the scenario being considered: 1) a UV-cutoff in the interaction Hamiltonian, that is, the probe no longer couples to high frequency modes, 2) a UV-cutoff in the field Hamiltonian, that is, the field no longer contains high frequency modes, and 3) a discrete approximation for the derivative. The effects of the first two changes on the probe's response to the field are exponentially suppressed with increasing $K$ and quickly become irrelevant for our calculations. The effect of the third change is more subtle, since the approximation that lies behind it is more drastic in nature, as becomes apparent by comparing the expression for the discretized derivative and the exact expression in Eq.~\eqref{derivative exact}. The discrete approximation for the derivative changes the dynamics of the field; namely, it modifies its dispersion relation from
\begin{align}
\hbar\omega_k=\sqrt{(m\,c^2)^2+(\hbar\,c\,k)^2}
\qquad\textrm{to}\qquad
\hbar\omega'_k&=\sqrt{(m\,c^2)^2+\left(\frac{2\,\hbar\,c\,K}{\pi}\right)^2\sin\left(\frac{\pi\, k}{2\, K}\right)^2}.
\end{align}
Note that $\hbar\omega_k\geq\hbar\omega'_k$, as seen in Fig. \ref{Dispersion}a). Note also that the dispersion relation is mostly modified at high frequencies, that is, at frequencies to which the probe does not couple strongly, as also shown in Fig.~\ref{Dispersion}a). One objection one may have, however, is that this modified dispersion relation allows for the possibility of superluminal signals to exist in these high frequency modes, but in practice, as we show, the probe does not couple to the field modes that behave pathologically.

To quantify how much the dispersion relation has changed \textit{in the modes that couple to the probe} we define the ``average relative error'' in $\hbar\omega_k$. This error is the average relative difference between the modified and unmodified dispersion relations at each frequency  weighted by the strength of the probe’s coupling to that frequency. That is,
\begin{align}
\text{Avg. Rel. Error} \coloneqq & \int_{-\infty}^\infty\d k \  \frac{\hbar\omega_k-\hbar\omega'_k}{\hbar\omega_k} \frac{\tilde{F}(k)}{\vert\!\vert \tilde{F}\vert\!\vert_1}.
\end{align}
We have computed the average relative error for various cutoffs, field masses, and probe sizes in Fig.~\ref{Dispersion}b). To investigate how the average relative error decreases as we increase $K$ we use the following series of inequalities,
\begin{align}
\text{Avg. Rel. Error}&=\!
\int_{-\infty}^\infty\!\!\!
\d k \  \frac{\hbar\omega_k-\hbar\omega'_k}{\hbar\omega_k} \ \frac{\tilde{F}(k)}{\vert\!\vert \tilde{F}\vert\!\vert_1}\leq\!
\int_{-\infty}^\infty\!\!\!
\d k \  \frac{(\hbar\omega_k)^2-(\hbar\omega'_k)^2}{(\hbar\omega_k)^2} \ \frac{\tilde{F}(k)}{\vert\!\vert \tilde{F}\vert\!\vert_1}\leq\!
\int_{-\infty}^\infty\!\!\!
\d k \  \frac{(\hbar\omega_k)^2-(\hbar\omega'_k)^2}{(\hbar\,c\,k)^2} \ \frac{\tilde{F}(k)}{\vert\!\vert \tilde{F}\vert\!\vert_1},
\end{align}
where the first inequality follows from $\hbar\omega_k \geq \hbar\omega'_k \geq 0$ and the second from $\hbar \omega_k \geq \hbar c k $. For the case in which the smearing function is a Gaussian of variance $\sigma^2$,
the final expression is mass-independent and can be computed in closed form, yielding
\begin{align}
\text{Avg. Rel. Error}&\leq 1 - 2\,\frac{K\sigma}{\sqrt{2\pi}}\text{Erf}\left(\frac{\pi}{K\sigma\sqrt{2}}\right)+\frac{2\,K^2\sigma^2}{\pi^2}\left(1-\exp\left(\frac{-\pi^2}{2\,K^2\sigma^2}\right)\right)\sim O\left(\frac{1}{(K\sigma)^2}\right).
\end{align}
That is, we expect the error made by the discrete approximation to be polynomially suppressed as we increase $K\sigma$. In the main text we have chosen $K\sigma=16$ in both the boundary sensing and thermometry examples. This places the upper-bound on average relative error at $0.32\%$. In the remote boundary sensing and thermometry examples we have (taking $\hbar=c=1$), $m\sigma=0.006$ and $m\sigma=0.00027$ respectively. The average relative error can be computed numerically in each case yielding $0.16\%$ in both cases.

\section{Total Variation and Hellinger Distances for Binary Boundary Classification}\label{appendixC}

In this appendix we add details on the analysis of the near-optimal performance of the neural network in our boundary sensing example (Sec.~\ref{BoundarySensing}). Notice that in this case study the probe remains Gaussian throughout its evolution under the Hamiltonians specified in the main text, as they are quadratic~\cite{Simon1988,Ferraro2005,Adesso2014}. Therefore, although it is not a necessary feature for our methods to work, in our first physical example the data is distributed normally and therefore the compression step is lossless. 

Consider the binary classification problem where we are asked to pick a label $y=0$ or $y=1$ for data $\bm{x}_0$ drawn from either $r_\theta(\bm{x})=p(\bm{x}|y=0,\theta)$ or from $q_\theta(\bm{x})=p(\bm{x}|y=1,\theta)$ with equal odds, where $\theta$ is some free parameter of the problem. In terms of the classification scenario considered in the main text, $\bm{x}$ is the local probe data, $y$ labels the field's boundary condition and $\theta$ describes the other details of the scenario, for instance $T_\textsc{min}$. The distributions $r_\theta(\bm{x})$ and $q_\theta(\bm{x})$ are then the odds that some particular data was produced given the state of the field and measurement procedure.

The optimal strategy (i.e., the one which maximizes your success probability) for this binary classification problem is to guess $y=1$ if \mbox{$q_\theta(\bm{x})>r_\theta(\bm{x})$} and $y=0$ if \mbox{$r_\theta(\bm{x})>q_\theta(\bm{x})$}, breaking ties randomly. This strategy succeeds with probability of $p_\text{success}=\frac{1}{2}\big(1+\text{TV}(r_\theta,q_\theta)\big)$ where 
\bel{TVdef}
\text{TV}(r_\theta,q_\theta)=\frac{1}{2}\int \big\vert r_\theta(\bm{x})-q_\theta(\bm{x})\big\vert \ \d\bm{x}
\ee
is the total variation distance between $r_\theta(\bm{x})$ and $q_\theta(\bm{x})$. If we can compute this distance, we can determine for which values of $\theta$ (e.g., for which coupling times) the distributions $r_\theta(\bm{x})$ and $q_\theta(\bm{x})$ are distinguishable. 

The total variation distance is only useful for binary classification problems (such as our remote boundary sensing example in the main text). It cannot be used in this way when there are more than two classes or for regression problems, such as our thermometry example. In the remote boundary sensing scenario, calculating the total variation distance directly is infeasible. An alternate approach is to compute upper and lower bounds on $TV$ using the Hellinger distance, $H(r_\theta,q_\theta)$ \cite{hellinger}, as
\bel{HellNormal}
H(r_\theta,q_\theta)^2
\leq\text{TV}(r_\theta,q_\theta)
\leq H(r_\theta,q_\theta)\sqrt{2-H(r_\theta,q_\theta)^2}
\quad\text{where}\quad
H(r,q)=\frac{1}{\sqrt{2}}\sqrt{\int\left(\sqrt{r(\bm{x})}-\sqrt{q(\bm{x})}\right)^2\d\bm{x}}.
\ee
Unfortunately, for the distribution of the compressed data, the Hellinger distance is not easier to compute. However, in the high tomography regime ($N_\textsc{tom}\gg1$) we can apply the central limit theorem to approximate $r_\theta$ and $q_\theta$ by multivariate normal distributions, $r_\theta(\bm{x})=\mathcal{N}(\bm{x};\mu_r,\Sigma_r)$ and $q_\theta(\bm{x})=\mathcal{N}(\bm{x};\mu_q,\Sigma_q)$ for some means, $\mu_r$ and $\mu_q$, and some covariances, $\Sigma_q$ and $\Sigma_r$. The Hellinger distance between two such multivariate normal distributions is given by \cite{hellinger}
\begin{align}
H(r_\theta,q_\theta)^2
&=1-\bigg(\frac{\text{det}(\Sigma_r\,\Sigma_q)}{\text{det}(\overline\Sigma^2)}\bigg)^\frac{1}{4}\!\!\exp\left(
-\frac{\Delta \mu^\intercal\overline\Sigma^{-1}\Delta\mu}{8}
\right),
\end{align}
where $\Delta\mu=\mu_r-\mu_q$ and $\overline\Sigma=(\Sigma_r+\Sigma_q)/2$. Thus if we can compute the means and covariances of our data in the central limit, we can find bounds for the neural network's optimal performance.

Due to the Gaussian nature of our setup all of our measurement results were drawn from normal distributions. Moreover, in the main text, we discussed how our data can be compressed by just considering the sample means and sample variances of each quadrature at each time point. For clarity we will restrict the following discussion to the results, $q_k$, of our $N_\textsc{tom}$ measurements of $\hat{q}$ at some \mbox{$t=T_\textsc{min}+m\Delta t$}, where \mbox{$m\in\{0,\dots,N_\textsc{times}-1\}$}. The sample mean and variance of these measurement outcomes are distributed as
\begin{align}
\bar{q}
&= \frac{1}{N_\textsc{tom}}\sum_{k=1}^{N_\textsc{tom}} q_k,
\sim \mathcal{N}\left(\mu_q,\frac{\sigma_{qq}}{N_\textsc{tom}}\right),
\quad\text{and}\quad
\bar{s}_q
= \frac{1}{N_\textsc{tom}} \sum_{k=1}^{N_\textsc{tom}} \big(q_k-\bar{q}\big)^2
\sim \sigma_{qq} \,  \frac{\chi^2(N_\textsc{tom}-1)}{N_\textsc{tom}},
\end{align}
where $\chi^2(k)$ is the chi-squared distribution with $k$ degrees of freedom \cite{cochran_1934} and $\mu_q=\langle \hat{q}\rangle$ and $\sigma_{qq}=\langle \hat{q}^2\rangle-\langle \hat{q}\rangle^2$ are the probe's first moment and variance in $\hat{q}$ at time $t$. Due to the Gaussian nature of our setup, these moments can be efficiently computed \cite{Schumaker,AdessoThesis}. Moreover, for independent identically distributed normal data, $q_k$, the sample mean and variance are sufficient statistics~\cite{Fisher1922} such that this compression is lossless.

Note that the compressed data is not normally distributed. However, for large $N_\textsc{tom}$, we can apply the central limit theorem yielding
\begin{align}
\bar{q}
\sim \mathcal{N}\left(\mu_q,\frac{\sigma_{qq}}{N_\textsc{tom}}\right),
\quad\text{and}\quad
\bar{s}_q
\sim \mathcal{N}\left(\sigma_{qq}, \frac{2\sigma_{qq}^2}{N_\textsc{tom}}\right). \end{align}
The same discussion applies equally well for our measurements of $\hat{p}$ and $\hat{r}$ at each time. Thus in the high tomography regime, our compressed data,
\begin{align}
\bm{x}=\left(\bar{q}(t),\bar{r}(t),\bar{p}(t),
\bar{s}_q(t),\bar{s}_r(t),\bar{s}_p(t)
\quad\text{for }t=T_\textsc{min}, \ T_\textsc{min}+\Delta t, \ \dots, \  T_\textsc{max}\right)
\end{align}
is distributed as $\bm{x}\sim\mathcal{N}(\bm{\mu},\Sigma)$ where
\begin{align}
   &\bm{\mu} = \left(\langle\hat{q}(t)\rangle,\langle\hat{r}(t)\rangle,\langle\hat{p}(t)\rangle,
    \sigma_{qq}(t),\sigma_{rr}(t),\sigma_{pp}(t)
    \quad\text{for }t=T_\textsc{min}, \ T_\textsc{min}+\Delta t, \ \dots, \  T_\textsc{max}\right)\\
   &\Sigma = \text{diag}\left(
   \dfrac{\sigma_{qq}(t)}{N_\textsc{tom}},
   \dfrac{\sigma_{rr}(t)}{N_\textsc{tom}},
   \dfrac{\sigma_{pp}(t)}{N_\textsc{tom}},
   \dfrac{2\sigma_{qq}^2(t)}{N_\textsc{tom}},
   \dfrac{2\sigma_{rr}^2(t)}{N_\textsc{tom}},
   \dfrac{2\sigma_{pp}^2(t)}{N_\textsc{tom}},
   \quad\text{for }t=T_\textsc{min}, \ T_\textsc{min}+\Delta t, \ \dots, \  T_\textsc{max}\right),
\end{align}
where $T_\textsc{max}=T_\textsc{min}+(N_\textsc{times}-1)\,\Delta t$. Knowing this distribution we can compute the Hellinger distance \eqref{HellNormal} and place bounds on optimal classification rate.

\section{How to calculate moments from non-Gaussian states}
\label{appendix4Moments}
In this appendix we show how to calculate the moments of the non-Gaussian state which results from applying a Gaussian unitary to a non-Gaussian state with known moments.

In particular, we will consider a Gaussian unitary interaction between a harmonic oscillator probe system, D, coupled to a quantum field, $\hat{\phi}(t,\bm{x})$, in a cavity. Let the probe be characterized by the dimensionless quadrature operators $\hat{q}_\textsc{d}$ and $\hat{p}_\textsc{d}$. The field operator can be decomposed into modes as
\begin{align}
\label{modedecompositionbis}
\hat{\phi}(t,\bm{x}) = 
\sqrt{\frac{2\hbar c^2}{L}}
\sum_{\ell=1}^\infty
\frac{\sin(k_\ell x)}{\sqrt{\omega_\ell}}
\big[\hat{q}_\ell\!\cos(\omega_\ell t)
+\hat{p}_\ell\!\sin(\omega_\ell t)\big],
\end{align}
where the dimensionless quadrature operators $\hat{q}_n$ and $\hat{p}_m$ satisfy canonical commutation relations  $[\hat{q}_n,\hat{p}_m]=\ii\delta_{nm}\openone$ and where $c\,k_\ell=\omega_\ell=\pi\,\ell\,c/L$. Let us collect all of these dimensionless quadrature operators together into a operator-valued phase space vector as
\begin{align}
\hat{\bm{X}}
\coloneqq
(\hat{q}_\textsc{d},\hat{p}_\textsc{d},\hat{q}_1,\hat{p}_1,\hat{q}_2,\hat{p}_2,\dots)^\intercal.    
\end{align}

\subsection{The effect of Gaussian unitaries on non-Gaussian states}
\label{effectGaussNonGauss}

Consider a scenario in which the initial state of the probe-field system is non-Gaussian with well known moments, e.g., the probe in its (Gaussian) ground state and the field in a Fock state. Suppose that the joint system undergoes a generic Gaussian unitary interaction, $\hat{U}$. Such a transformation maps Gaussian states to Gaussian states. In this case there is a well known relationship between the first and second moments of the initial and final states. Conversely, when applied to a non-Gaussian state, the result of such a transformation would be another non-Gaussian state. Note that in this case, since the field is in a non-Gaussian state, the reduced state of the probe may end up being non-Gaussian even if its initial state was Gaussian.

However, despite this complication, just as in the Gaussian case, there is still a relatively simple systematic relationship between all of the higher-moments before and after this transformation. That is, given the moments of the initial non-Gaussian state, we can efficiently determine all of the moments of the final non-Gaussian state.

To show this, we first note that Gaussian unitary transformations correspond to symplectic transformations in phase space. That is, if we act on each component of the vector of operators, $\hat{\bm{X}}$, via some Gaussian unitary, $\hat{U}$, the result is a symplectic(-affine) transformation of the phase space vector $\hat{\bm{X}}$ itself. Namely,
\begin{align}
\hat{U}^\dagger\hat{\bm{X}}\hat{U}
\coloneqq
\begin{pmatrix}
\hat{U}^\dagger\hat{X}_1\hat{U}\\
\hat{U}^\dagger\hat{X}_2\hat{U}\\
\hat{U}^\dagger\hat{X}_3\hat{U}\\
\hat{U}^\dagger\hat{X}_4\hat{U}\\
\dots
\end{pmatrix}
=S\begin{pmatrix}
\hat{X}_1\\
\hat{X}_2\\
\hat{X}_3\\
\hat{X}_4\\
\dots
\end{pmatrix}
+\begin{pmatrix}
d_1\hat{\openone}\\
d_2\hat{\openone}\\
d_3\hat{\openone}\\
d_4\hat{\openone}\\
\dots
\end{pmatrix}
\eqqcolon S\,\hat{\bm{X}}+\bm{d}\hat{\openone}
\end{align}
for some symplectic transformation $S$ with $S\Omega S^\intercal=\Omega$ (where $\Omega$ is the symplectic form) and some displacement vector $\bm{d}$. Note that in the above equation $\hat{U}$ acts as a linear map on the system's Hilbert space and acts on $\bm{\hat{X}}$ component-wise. On the other hand, $S$ is a linear map on the system's phase space and acts on $\bm{\hat{X}}$ as a phase space vector, yielding linear combinations of its (operator-valued) components.

We note that this Gaussian unitary (or the equivalent symplectic) relationship is a property of the dynamics alone, independent of the system state. Thus, we can use this characterization of Gaussian unitaries to understand their effect on non-Gaussian states. The effect that a Gaussian unitary has on a state's Wigner function (even a non-Gaussian one) is,
\begin{align}\label{WignerUpdate}
U:\quad W_{\hat{\rho}}(\bm{\xi})\to W_{\hat{U}\hat{\rho}\hat{U}^\dagger}(\bm{\xi})= W_{\hat{\rho}}(S^{-1}\bm{\xi}-\bm{d}).
\end{align}
That is, the effect of a Gaussian unitary is just to transform the original Wigner function by applying a linear-affine transformation to the joint phase space variables. Note that $\bm{\xi}$ in the above equation is the real-valued vector of phase space variables, $\bm{\xi}\coloneqq
(q_\textsc{d},p_\textsc{d},q_1,p_1,q_2,p_2,\dots)^\intercal$. For simplicity we will now restrict our attention to situations with zero displacement, $\bm{d}=0$, since this is the case relevant for this paper.

Next, we can use \eqref{WignerUpdate} to determine the moments of the final probe distribution from the initial probe field moments. For instance, suppose that we are interested in the fourth moment of the probe's quadrature $\hat{q}_\textsc{d}$ after the interaction with the field. We can calculate this as follows. Let $\bm{q}_\textsc{d}=(1,0,0,\dots)^\intercal$ be a phase space vector such that $\bm{q}_\textsc{d}^\intercal\bm{\xi}=q_\textsc{d}$. That is, $\bm{q}_\textsc{d}$ isolates $q_\textsc{d}$ from the vector of phase space variables $\bm{\xi}$. This allows us to rewrite the desired fourth moment as
\begin{align}
\langle \hat{q}_\textsc{d}^4\rangle_{\hat{U}\hat{\rho}\hat{U}}
&=\int \d\bm{\xi} \ 
q_\textsc{d}^4 \ 
W_{\hat{U}\hat{\rho}\hat{U}}(\bm{\xi})
=\int \d\bm{\xi} \ 
(\bm{q}_\textsc{d}^\intercal\bm{\xi})^4 \ 
W_{\hat{\rho}}(S^{-1}\bm{\xi}).
\end{align}
Note we have assumed $\bm{d}=\bm 0$ for simplicity. Making a canonical change of variables to $\bm{\xi'}=S^{-1}\bm{\xi}$ we have
\begin{align}
\langle \hat{q}_\textsc{d}^4\rangle_{\hat{U}\hat{\rho}\hat{U}}
&=\int \d\bm{\xi'} \ 
(\bm{q}_\textsc{d}^\intercal S\bm{\xi'})^4 \
W_{\hat{\rho}}(\bm{\xi'})
=\int \d\bm{\xi'} \ 
(\bm{Q}_\textsc{d}^\intercal \bm{\xi'})^4 \
W_{\hat{\rho}}(\bm{\xi'}) \;,
\end{align}
where we have defined $\bm{Q}_\textsc{d}\coloneqq S^{\intercal}\bm{q}_\textsc{d}\Rightarrow \bm{q}_\textsc{d}=S^{-\intercal} \bm{Q}_\textsc{d} $. Note that for all symplectic transformations,  $\det(S)=1$, such that no Jabobian factor arises in the above change of variables.  The fourth moment of $q_\textsc{d}$ in the final state is equal to the fourth moment of $\bm{Q}_\textsc{d}$ in the initial state. Since we have assumed that we know all of the moments of the initial non-Gaussian state, we can, at least in principle, calculate $\langle \hat{q}_\textsc{d}^4\rangle_{\hat{U}\hat{\rho}\hat{U}}$. In general,  $\bm{Q}_\textsc{d}$ will have support over  the probe portion of the phase spaces as well as over a great deal of the field phase space. As such, calculating $\langle \hat{q}_\textsc{d}^4\rangle_{\hat{U}\hat{\rho}\hat{U}}$ requires us to know the correlations between all of the field modes. In particular we have
\begin{align}
\langle \hat{q}_\textsc{d}^4\rangle_{\hat{U}\hat{\rho}\hat{U}}
&=\!\int\! \d\bm{\xi} \ 
(\bm{Q}_\textsc{d}^\intercal \bm{\xi})^4 \
W_{\hat{\rho}}(\bm{\xi}) = \!\!\int\! \d\bm{\xi}
\bigg(\sum_{\ell=1}^\infty Q_{\textsc{d},\ell} \, \xi_\ell\bigg)^4 \ 
W_{\hat\rho}(\bm{\xi}) = \!\int\! \d\bm{\xi}
\sum_{i,j,k,\ell=1}^\infty
Q_{\textsc{d},i} \, Q_{\textsc{d},j}  \, Q_{\textsc{d},k} \, Q_{\textsc{d},\ell} \, 
\xi_i \, 
\xi_j \,
\xi_k \, 
\xi_\ell \, 
W_{\hat\rho}(\bm{\xi}) \\
&=\sum_{i,j,k,\ell=1}^\infty
Q_{\textsc{d},i} \, Q_{\textsc{d},j}  \, Q_{\textsc{d},k} \, Q_{\textsc{d},\ell} \, 
\int \d\bm{\xi} \ 
\xi_i \, 
\xi_j \,
\xi_k \, 
\xi_\ell \, 
W_{\hat\rho}(\bm{\xi})  = \sum_{i,j,k,\ell=1}^\infty
Q_{\textsc{d},i} \, Q_{\textsc{d},j}\, Q_{\textsc{d},k} \, Q_{\textsc{d},\ell} \, 
T_{4,i,j,k,\ell} \nonumber\\
&=T_4(\bm{Q}_\textsc{d},\bm{Q}_\textsc{d},\bm{Q}_\textsc{d},\bm{Q}_\textsc{d}) \nonumber 
\end{align}
where $T_{4,i,j,k,\ell}=\int \d\bm{\xi} \  \xi_i \,  \xi_j \, \xi_k \,  \xi_\ell \,  W_{\hat\rho}(\bm{\xi})$ are the fourth moments of the initial Wigner function, and $T_4$ collects these. The desired probe moment, $\langle \hat{q}_\textsc{d}^4\rangle_{\hat{U}\hat{\rho}\hat{U}}$, is this tensor evaluated on four copies of the phase space vector $\bm{Q}_\textsc{d}$. Similarly, for the fourth moments of $\hat{p}_\textsc{d}$ and $\hat{r}_\textsc{d}$,
\begin{align}
\langle \hat{p}_\textsc{d}^4\rangle_{\hat{U}\hat{\rho}\hat{U}} 
&=T_4(\bm{P}_\textsc{d},\bm{P}_\textsc{d},\bm{P}_\textsc{d},\bm{P}_\textsc{d}),\\
\langle \hat{r}_\textsc{d}^4\rangle_{\hat{U}\hat{\rho}\hat{U}} 
&=T_4(\bm{R}_\textsc{d},\bm{R}_\textsc{d},\bm{R}_\textsc{d},\bm{R}_\textsc{d}),
\end{align}
where $\bm{p}_\textsc{d}\coloneqq(0,1,0,\dots)^\intercal$ and $\bm{P}_\textsc{d}\coloneqq S^{\intercal}\bm{p}_\textsc{d}$, $\bm{R}_\textsc{d}=(\bm{Q}_\textsc{d}+\bm{P}_\textsc{d})/\sqrt{2}$. Any other final probe moments (e.g. $\langle \hat{q}_\textsc{d}^2\rangle_{\hat{U}\hat{\rho}\hat{U}}$ or $\langle \hat{p}_\textsc{d}^8\rangle_{\hat{U}\hat{\rho}\hat{U}}$) can be calculated in analogous ways. All that one needs is an understanding of the moments of the initial non-Gaussian state (i.e., the tensors $T_2$, $T_4$, $T_8$, etc.) as well as the vectors $\bm{Q}_\textsc{d}$ and $\bm{P}_\textsc{d}$ which the Gaussian unitary interaction maps into the probe's phase space.

\subsection{Two delta pulse interaction}
As we showed in the Subsec.~\ref{effectGaussNonGauss}, in order to construct the desired probe moments from the moments of the initial non-Gaussian state, we need to know the vectors $\bm{Q}_\textsc{d}\coloneqq S^{\intercal}\bm{q}_\textsc{d}$ and  $\bm{P}_\textsc{d}\coloneqq S^{\intercal}\bm{p}_\textsc{d}$. These are the vectors which the interaction maps onto the probe observables $\hat{q}_\textsc{d}$ and $\hat{p}_\textsc{d}$. To identify these vectors we first need to compute the symplectic transformation $S$ which is associated with our Gaussian unitary evolution.

Let us consider an interaction Hamiltonian given by \eqref{Hint0} with a switching function which is the sum of two delta functions, specifically, $\chi(t)=\delta(t)+\delta(t-t_\textsc{m})$. In this scenario the probe undergoes a very strong, very brief interaction with the field once at $t=0$ and then once again at $t=t_\textsc{m}>0$. Between these two times, the probe evolves freely. As such, the operator with which it couples to the field, $\hat{\mu}_\textsc{d} =\hat{q}_\textsc{d}$, evolves in the interaction picture as
\begin{equation}
   \hat{\mu}_\textsc{d}(t) = \hat{q}_\textsc{d}(t)= \hat{q}_\textsc{d}\cos(\omega_\textsc{d} t)
+\hat{p}_\textsc{d}\sin(\omega_\textsc{d} t)
\end{equation}
where $\omega_\textsc{d}$ is the probe's natural frequency.
The full unitary map resulting from both of these sudden interactions is given by
\begin{align}
\hat{U} = \hat{U}_\text{int}(t_\textsc{m}) \ \hat{U}_\text{int}(0),
\end{align}
where
\begin{align}
\hat{U}_\text{int}(t)&=\exp\left(-\ii\lambda \ 
\hat{\mu}_\textsc{d}(t)
\otimes
\int\d x \, F(x) \,  \hat{\phi}(t,x)
\right).
\end{align}
Rewriting this in terms of the operator valued phase space vector, $\hat{\bm{X}}=
(\hat{q}_\textsc{d},\hat{p}_\textsc{d},\hat{q}_1,\hat{p}_1,\hat{q}_2,\hat{p}_2,\dots)^\intercal$, we have 
\begin{align}
\hat{U}_\text{int}(t)
&=\exp\left(-\ii\lambda \ 
\frac{1}{2}\bm{\hat{X}}^\intercal \mathcal{H}(t) \bm{\hat{X}}
\right),
\end{align}
where $\mathcal{H}(t)$ is the following bilinear form:
\begin{align}
\mathcal{H}(t) 
&= \bm{u}(t) \bm{v}(t)^\intercal+\bm{v}(t) \bm{u}(t)^\intercal,\\
\bm{u}(t)
&=(\cos(\omega_\textsc{d} t),\sin(\omega_\textsc{d} t),0,0,\dots)^\intercal,\\
\bm{v}(t)
&=\bigg(0,0,
\frac{F_1\cos(\omega_1 t)}{\sqrt{\omega_1}},
\frac{F_1\sin(\omega_1 t)}{\sqrt{\omega_1}},
\frac{F_2\cos(\omega_2 t)}{\sqrt{\omega_2}},
\frac{F_2\sin(\omega_2 t)}{\sqrt{\omega_2}},
\dots\bigg)^\intercal,
\end{align}
where \mbox{$F_n=\sqrt{2\hbar c^2/L}\int\d x F(x)\sin\left(k_n x\right)$}. Note that $\bm{u}(t)$ has support only on the probe sector of the phase space. It tracks the evolution of the probe observable, $\hat{\mu}_\textsc{d} =\hat{q}_\textsc{d}$, through time. Similarly, $\bm{v}(t)$ has support only on the field sector of the phase space, and it tracks the evolution of the field observable, $\int\d \bm{x} \, F(\bm{x}) \,  \hat{\phi}(t,\bm{x})$.

The symplectic transformation $S$ corresponding to $U$ is given by
\begin{align}
S=S_\text{int}(t_\textsc{m})\,S_\text{int}(0),    
\end{align}
where
\begin{align}
S_\text{int}(t)=\text{exp}(\lambda\,\Omega\,\mathcal{H}(t)).    
\end{align}
In this case, the above symplectic transformations are easy to compute. This is due to the fact that the matrix $\Omega\mathcal{H}(t)$ is nilpotent, in particular, $(\Omega\mathcal{H}(t))^2=0$. This follows from the four orthogonality relations
\begin{align}
\bm{v}(t)^\intercal\Omega\bm{u}(t)=0,
\quad
\bm{u}(t)^\intercal\Omega\bm{u}(t)=0,
\quad
\bm{v}(t)^\intercal\Omega\bm{v}(t)=0,
\quad
\bm{u}(t)^\intercal\Omega\bm{v}(t)=0.
\end{align}
The four equalities hold because $\Omega$ is anti-symmetric and because $\Omega$ does not mix the probe and field portions of the phase space (therefore, $\Omega\bm{u}(t)$ and $\bm{v}(t)$ have no common support). The nilpotence of $\Omega\mathcal{H}(t)$ makes the matrix exponential trivial to compute. Indeed since $\Omega\mathcal{H}(t)$ squares to the zero operator we have
\begin{align}
S_\text{int}(t)&=\openone+\lambda\Omega\mathcal{H}(t)
=\openone+\lambda \, \Omega\,(\bm{u}(t)\bm{v}(t)^\intercal
+\bm{v}(t)\bm{u}(t)^\intercal),\\
S_\text{int}^{-1}(t)&=\openone-\lambda\Omega\mathcal{H}(t)
=\openone-\lambda \, \Omega\,(\bm{u}(t)\bm{v}(t)^\intercal
+\bm{v}(t)\bm{u}(t)^\intercal).
\end{align}
Note that these expressions are exact, not perturbative. Applying these transformations to $\bm{q}_\textsc{d}\coloneqq(1,0,0,\dots)^\intercal$ and $\bm{p}_\textsc{d}\coloneqq(0,1,0,\dots)^\intercal$ gives $\bm{Q}_\textsc{d}$ and $\bm{P}_\textsc{d}$ as desired.

\section{Second, fourth and eighth moments for Fock and Phase-averaged states}\label{appendifxFockCoherent}

In this Appendix we will calculate the second, fourth and eighth moments for Fock and Phase-averaged states. We consider the initial probe-field Wigner function defined in Eq.~\eqref{fockAndcohe} in the main text,
\begin{align}
W(q_\textsc{d},p_\textsc{d},q_1,p_1,q_2,p_2,\dots ;N)&=
W_\text{Vac}(q_\textsc{d},p_\textsc{d})
\times
\left(W_\text{Fock}(q_1,p_1;N)\text{ or }
W_\text{PAC}(q_1,p_1;N)\right)
\times \Pi_{n=2}^\infty \ W_\text{Vac}(q_n,p_n) \;,
\end{align}
where $q_\textsc{d}$ and $p_\textsc{d}$ are the probe variables, and where $W_\text{Vac}(q,p)
= e^{-q^2-p^2}/\pi$. Note that the probe and all the modes are uncorrelated from each other. Thus, all ``cross moments'' factorize (e.g.,  $\langle q_4 p_4^2 q_6^3p_6^5\rangle=\langle q_4 p_4^2\rangle \langle q_6^3p_6^5\rangle$). Note also that these averages are taken with respect to the Wigner function, so that when we calculate the average of a certain dynamical function $f(\bm{\xi})$, which we will denote $\langle f(\bm{\xi})\rangle$, this corresponds to the expectation value of the Weyl quantized operator $\langle \hat f(\bm{\xi}) \rangle$, which is the one that we obtain by using symmetric ordering in the quantization scheme~\cite{Weyl1927,Wigner1932,Hillery1984}. 

The statistics of the vacuum Wigner function are
\begin{align}
\langle q^2\rangle_\text{Vac}=1/2,
\quad
\langle q^4\rangle_\text{Vac}=3/4,
\quad
\langle q^6\rangle_\text{Vac}=15/8,
\quad
\langle q^8\rangle_\text{Vac}=105/16,
\end{align}
with $\langle p^n\rangle=\langle q^n\rangle$, $\langle p^n q^m\rangle=\langle p^n\rangle \langle q^m\rangle$, and odd moments vanishing. We only write explicitly up to the eighth moments that we need for our purposes, but of course for the modes in the vacuum  all their odd moments vanish and their even moments are trivial functions of their second moments.

The statistics ($\langle q^2\rangle_N, \ \langle q^4\rangle_N, \ \langle q^6\rangle_N, \ \langle q^8\rangle_N$) of the Fock and PAC states Wigner functions are given in Tables~\ref{FockN} and~\ref{CoherentN}, respectively, with \mbox{$\langle p^n\rangle=\langle q^n\rangle$}, \mbox{$\langle p^n q^m\rangle=\langle p^n\rangle \langle q^m\rangle$}, and the odd moments vanishing. Note that second moments match (first column). Also, the first row is the same for both tables since for $N=0$ both the Fock state and the PAC reduce to the vacuum.
\begin{table}[ht]
    \centering
    \begin{tabular}{|c|cccc|}
         \hline\ &  \ $\langle q^2\rangle_N$  \ &  \ $\langle q^4\rangle_N$ \  &  \ $\langle q^6\rangle_N$  \ &  \ $\langle q^8\rangle_N$  \ \\
         \hline
         N=0 & 1/2 & 3/4 & 15/8 & 105/16\\
         N=1 & 3/2 & 15/4 & 105/8 & 945/16\\
         N=2 & 5/2 & 39/4 & 375/8 & 4305/16\\
         N=3 & 7/2 & 75/4 & 945/8 & 13545/16\\
         N=4 & 9/2 & 123/4 & 1935/8 & 33705/16\\
         \hline
    \end{tabular}
    \caption{Statistics for the $N$ particle Fock states}\label{FockN}
\end{table}

\begin{table}[ht]
    \centering
    \begin{tabular}{|c|cccc|}
         \hline\ &  \ $\langle q^2\rangle_N$  \ &  \ $\langle q^4\rangle_N$ \  &  \ $\langle q^6\rangle_N$  \ &  \ $\langle q^8\rangle_N$  \ \\
         \hline
         N=0 & 1/2 & 3/4 & 15/8 & 105/16\\
         N=1 & 3/2 & 21/4 & 215/8 & 2835/16\\
         N=2 & 5/2 & 51/4 & 715/8 & 12425/16\\
         N=3 & 7/2 & 93/4 & 1635/8 & 34755/16\\
         N=4 & 9/2 & 147/4 & 3095/8 & 77385/16\\
         \hline
    \end{tabular}
    \caption{Statistics for the phase averaged coherent state with $\langle\hat{n}\rangle=N$ particle on average}\label{CoherentN}
\end{table}

In Appendix \ref{appendix4Moments}, we obtained a general formula to calculate the moments of the probe's observables for a probe delta-coupled  to a general field at times $t = 0$ and $t =t_\textsc{m}$. We considered a harmonic oscillator probe, $\hat{\mu}_\textsc{d} = \hat{q}_\textsc{d}$, with frequency $\omega_\textsc{d}$. In the next subsection, we will apply those results to the particular cases of Tables~\ref{FockN} and~\ref{CoherentN}. 

\subsection{Second moments}

 Using the techniques in Appendix \ref{appendix4Moments}, the general formula for $\langle \hat{q}_\textsc{d}(t)^2\rangle $, $\langle \hat{p}_\textsc{d}(t)^2\rangle $ and $\langle \hat{r}_\textsc{d}(t)^2\rangle $ for $t> t_\textsc{m}$ can be obtained as
 \begin{align}
     \langle \hat{q}_\textsc{d}(t)^2\rangle  &= \bm{Q}_\textsc{d}^\intercal
\langle \hat{\bm{X}} \hat{\bm{X}}^\intercal \rangle
\bm{Q}_\textsc{d}=\bm{Q}_\textsc{d}^\intercal
\sigma_0
\bm{Q}_\textsc{d}\,, \label{secondmomentqD}\\
\langle \hat{p}_\textsc{d}(t)^2\rangle  &= \bm{P}_\textsc{d}^\intercal
\langle \hat{\bm{X}} \hat{\bm{X}}^\intercal \rangle
\bm{P}=\bm{P}_\textsc{d}^\intercal
\sigma_0
\bm{P}\textsc{d}\,, \label{secondmomentpD}\\
\langle \hat{r}_\textsc{d}(t)^2\rangle  &= \frac{1}{2}(\bm{Q}_\textsc{d}+\bm{P}_\textsc{d})^\intercal
\langle \hat{\bm{X}} \hat{\bm{X}}^\intercal \rangle
(\bm{Q}_\textsc{d}+\bm{P}_\textsc{d})=\frac{1}{2}(\bm{Q}_\textsc{d}+\bm{P}_\textsc{d})^\intercal
\sigma_0
(\bm{Q}_\textsc{d}+\bm{P}_\textsc{d})\,, \label{secondmomentrD}
 \end{align}
 where $\bm{Q}_\textsc{d}=S^{\intercal}\bm{q}_\textsc{d}$ and  $\bm{P}_\textsc{d}=S^{\intercal}\bm{p}_\textsc{d}$ can be calculated as
 \begin{align}
     \bm{Q}_\textsc{d}&= S^{\intercal}\bm{q}_\textsc{d}=S^{\intercal}_{\text{int}}(0) S^{\intercal}_{\text{int}}(t_{\textsc{m}})\bm{q}_\textsc{d} 
     =(\openone-\lambda(\bm{u}(0)\bm{v}^\intercal(0)+ \bm{v}(0)\bm{u}^\intercal(0))\Omega)(\openone- \lambda(\bm{u}(t_{\textsc{m}})\bm{v}^\intercal(t_{\textsc{m}})+ \bm{v}(t_{\textsc{m}})\bm{u}^\intercal(t_{\textsc{m}}))\Omega)\bm{q}_\textsc{d} \label{QDtM}\\
     &=(1-\lambda^2 \beta \sin(\omega_\textsc{d}t_\textsc{m}))\bm{q}_\textsc{d}+\lambda \sin(\omega_\textsc{d}t_\textsc{m})\bm{v}(t_\textsc{m}) \;, \nonumber\\
    \bm{P}_\textsc{d}&= S^{\intercal}\bm{p}_\textsc{d}=S^{\intercal}_{\text{int}}(0)S^{\intercal}_{\text{int}}(t_\textsc{m})\bm{p}_\textsc{d} = (\openone-\lambda(\bm{u}(0)\bm{v}^\intercal(0)+ \bm{v}(0)\bm{u}^\intercal(0))\Omega)(\openone- \lambda(\bm{u}(t_{\textsc{m}})\bm{v}^\intercal(t_{\textsc{m}})+ \bm{v}(t_{\textsc{m}})\bm{u}^\intercal(t_{\textsc{m}}))\Omega)\bm{p}_\textsc{d} \label{PDtM}\\
     &=\bm{p}_\textsc{d}+\lambda^2 \beta \cos(\omega_\textsc{d}t_\textsc{m}) \bm{q}_\textsc{d}-\lambda(\bm{v}(0)+\cos(\omega_\textsc{d}t_\textsc{m})\bm{v}(t_\textsc{m})) \;, \nonumber
 \end{align}
 where
\begin{align}
&\bm{u}(t)=(\cos(\omega_\textsc{d} t),\sin(\omega_\textsc{d} t),0,0,\dots)^\intercal,\\
&\bm{v}(t)=\bigg(0,0,
\frac{F_1\cos(\omega_1 t)}{\sqrt{\omega_1}},
\frac{F_1\sin(\omega_1 t)}{\sqrt{\omega_1}},
\frac{F_2\cos(\omega_2 t)}{\sqrt{\omega_2}},
\frac{F_2\sin(\omega_2 t)}{\sqrt{\omega_2}},
\dots\bigg)^\intercal, \\
&\beta=\bm{v}(0)^\intercal\Omega\bm{v}(t_\textsc{m})=\sum_{n=1}^\infty \frac{F_n^2}{\omega_n}\sin(\omega_n t_\textsc{m})\,, \label{betaformula} \\
&F_n=\sqrt{2\hbar c^2/L}\int\d x F(x)\sin(k_n x)\,.
\end{align}
In Eqs.~\eqref{secondmomentqD}---\eqref{betaformula} we observe that the only dependence on the initial state of the field is encoded in $\sigma_0$, the covariance matrix of the initial states. Therefore, the only further particularization that we need to perform is to substitute the covariance matrix of the initial field state for each particular case, which can be computed using the values in Tables~\ref{FockN} and~\ref{CoherentN}. 

\subsection{Fourth moments}

From Appendix~\ref{appendix4Moments} we know that 
\begin{align}
    \langle \hat{q}_\textsc{d}(t)^4\rangle &=\sum_{i,j,k,\ell=1}^\infty
Q_i(t) \, Q_j(t)  \, Q_k(t) \, Q_\ell(t) \, 
T_{4,i,j,k,\ell} \,,\\
    \langle \hat{p}_\textsc{d}(t)^4\rangle &=\sum_{i,j,k,\ell=1}^\infty
P_i(t) \, P_j(t)  \, P_k(t) \, P_\ell(t) \, 
T_{4,i,j,k,\ell}\,,\\
\langle \hat{r}_\textsc{d}(t)^4\rangle &=\sum_{i,j,k,\ell=1}^\infty
\frac{1}{4}(Q_i(t) + P_i(t))\, (Q_j(t) + P_j(t))  \, (Q_k(t) + P_k(t)) \, (Q_l(t) + P_l(t)) \, 
T_{4,i,j,k,\ell} \,,
\end{align}
where $T_{4,i,j,k,\ell}=\int \d\bm{\xi} \,
\xi_i \, 
\xi_j \,
\xi_k \, 
\xi_\ell \, 
W(\bm{\xi}) =\langle \xi_i \, 
\xi_j \,
\xi_k \, 
\xi_\ell \rangle$. We observe that $T_{4,i,j,k,\ell}$ is fully symmetric under index permutation, i.e., \mbox{$T_{4,i,j,k,\ell} =T_{4,\tau(i),\tau(j),\tau(k),\tau(\ell)} $} for any 4-permutation $\tau$. We also observe that only the terms of the form $T_{4,i,i,i,i}$ or \mbox{$T_{4,i,i,j,j} = T_{4,i,j,j,i}= T_{4,i,j,i,j}= T_{2,i,i}T_{2,j,j}$} for $i\neq j$ are different from zero. Taking into account the different permutations, we obtain: 
\begin{align}
  \langle \hat{q}_\textsc{d}(t)^4\rangle  &=\sum_{i,j,k,\ell=1}^\infty
Q_i(t) \, Q_j(t)  \, Q_k(t) \, Q_\ell(t) \, 
T_{4,i,j,k,\ell} \\
\nonumber
&=  \sum_{i=1}^\infty
Q_i^4(t) T_{4,i,i,i,i} + 3\!\!\!\!\!\!\!\!\sum_{i=1, j=1, j\neq i}^\infty
Q_i(t)^2\, Q_j(t)^2 \,T_{2,i,i}T_{2,j,j} + 3 \!\!\!\sum_{i=1, j=i}^\infty
Q_i^4(t) T_{2,i,i}T_{2,i,i} - 3 \!\!\!\sum_{i=1, j=i}^\infty
Q_i^4(t) T_{2,i,i}^2 \\
&= \sum_{i=1}^\infty
Q_i^4(t) (T_{4,i,i,i,i}-3 T_{2,i,i}^2)+ 3\bigg(\sum_{i=1}^\infty
Q_i(t)^2\,T_{2,i,i}\bigg)^2 \;. \nonumber
\end{align}
We additionally have that, if $i\neq 3, 4$, then $T_{4,i,i,i,i}-3 T_{2,i,i}^2 = \frac{3}{4} - 3 \big(\frac{1}{2}\big)^2 = 0$. Therefore,
\begin{equation}
    T_{4,i,i,i,i}-3 T_{2,i,i}^2= \bigg\{
  \begin{array}{lr} 
      0 & i\neq 3,4\\
      \langle q^4 \rangle_N-3\langle q^2 \rangle_N^2  & i=3,4
      \end{array}
\end{equation}
and
\begin{equation}
    T_{2,i,i}= \bigg\{
  \begin{array}{lr} 
      1/2 & i\neq 3,4\\
     \langle q^2 \rangle_N  & i=3,4
      \end{array} \,.
\end{equation}
Proceeding analogously we get (with $R = (Q+P)/\sqrt{2}$) 
\begin{align}
    \langle \hat{p}_\textsc{d}(t)^4\rangle &=\sum_{i=1}^\infty
P_i^4(t) (T_{4,i,i,i,i}-3 T_{2,i,i}^2)+ 3\bigg(\sum_{i=1}^\infty
P_i(t)^2\,T_{2,i,i}\bigg)^2 \,, \\
\langle \hat{r}_\textsc{d}(t)^4\rangle &= \sum_{i=1}^\infty
R_i^4(t) (T_{4,i,i,i,i}-3 T_{2,i,i}^2)+ 3\bigg(\sum_{i=1}^\infty
R_i(t)^2\,T_{2,i,i}\bigg)^2 \;.
\end{align}

\subsection{Eighth moments}
In order to calculate the eight moments, as we did with the fourth moments, we can take advantage of the symmetries of the tensor $T_{8,i,j,k,l,m,n,o,p}$. As stated before, $T_{8,i,j,k,l,m,n,o,p}=T_{8,\tau(i),\tau(j),\tau(k),\tau(l),\tau(m),\tau(n),\tau(o),\tau(p)}$ for any 8-permutation $\tau$. As a consequence, several terms cancel out, yielding:
\begin{align}
\langle \hat{q}_\textsc{d}(t)^8\rangle &=\sum_{i,j,k,\ell, m,n,o,p=1}^\infty
Q_i(t) \, Q_j(t)  \, Q_k(t) \, Q_\ell(t) \,Q_m(t) \, Q_n(t) \, Q_o(t) \, Q_p(t) \,  
T_{8,i,j,k,\ell,m,n,o,p} \\\nonumber
&=\sum_{i=1}^\infty T_{8,i,i,i,i,i,i,i,i}Q_i^8(t)+28\sum_{i,j=1, j\neq i}^\infty T_{8,i,i,i,i,i,i,j,j}Q_i^6(t)Q_j^2(t)\\
\nonumber
&\phantom{=,}+35\sum_{i,j=1, j\neq i}^\infty T_{8,i,i,i,i,j,j,j,j}Q_i^4(t)Q_j^4(t) + 210 \sum_{i,j,k=1, j\neq i, k\neq i,k\neq j}^\infty T_{8,i,i,i,i,j,j,k,k}Q_i^4(t)Q_j^2(t)Q_k^2(t) \\
&\phantom{=,}+105\sum_{i,j,k,l=1, j\neq i, k\neq i,j ,l\neq i,j,k}^\infty T_{8,i,i,j,j,k,k,l,l}Q_i^2(t)Q_j^2(t)Q_k^2(t)Q_l^2(t) \;. \nonumber
\end{align}
Now let us look at the terms one by one. The second term's coefficient comes from the possible combinations with six equal indices and two equal indices (these two different from the previous six). Therefore we have to choose six indices from the eight available to be equal to $i$ and then two from the remaining two to set to $j$: ${8 \choose 6}{2 \choose 2} = 28$. Then, we have to sum over all the pairs (and the order of the pairs matters). Now, since $T_{8,i,i,i,i,i,i,j,j}=T_{6,i,i,i,i,i,i}T_{2,j,j}$, we get
\begin{align}
  28\sum_{i,j=1, j\neq i}^\infty T_{6,i,i,i,i,i,i}T_{2,j,j}Q_i^6(t)Q_j^2(t)  &= 28\sum_{i,j=1}^\infty T_{6,i,i,i,i,i,i}T_{2,j,j}Q_i^6(t)Q_j^2(t) - 28\sum_{i=1}^\infty T_{6,i,i,i,i,i,i}T_{2,i,i}Q_i^8(t) \\\nonumber
  &=28 \bigg(\sum_{i=1}^\infty T_{6,i,i,i,i,i,i} Q_i^6(t)\bigg)\bigg(\sum_{j=1}^\infty T_{2,j,j}Q_j^2(t)\bigg) - 28\sum_{i=1}^\infty T_{6,i,i,i,i,i,i}T_{2,i,i}Q_i^8(t) \;.
\end{align}
The third term's coefficient comes from the possible combinations of two quadruples of equal indices different between each other. Therefore, we have to choose four indices from the eight available to be equal to $i$ and then four from the remaining four to set to $j$: ${8 \choose 4}{4 \choose 4} = 70$. Then, we have to sum over all the ordered pairs, but since in this case order does not matter, we have to divide by two. Now, since $T_{8,i,i,i,i,j,j,j,j}=T_{4,i,i,i,i}T_{4,j,j,j,j}$, we get
\begin{align}
  35\sum_{i,j=1, j\neq i}^\infty T_{4,i,i,i,i}T_{4,j,j,j,j}Q_i^4(t)Q_j^4(t)  &= 35\sum_{i,j=1}^\infty T_{4,i,i,i,i}T_{4,j,j,j,j}Q_i^4(t)Q_j^4(t) - 35\sum_{i=1}^\infty T_{4,i,i,i,i}^2 Q_i^8(t) \\
  &=35 \bigg(\sum_{i=1}^\infty T_{4,i,i,i,i} Q_i^4(t)\bigg)^2 - 35\sum_{i=1}^\infty T_{4,i,i,i,i}^2 Q_i^8(t) \;. \nonumber
\end{align}
The fourth term's factor comes from the possible combinations of a quadruple and two pairs of equal indices, different between each other. Therefore, we have to choose four indices from the eight available to be equal to $i$, then two from the remaining four to set to $j$ and then the last two to set to $k$ : ${8 \choose 4}{4 \choose 2}{2 \choose 2} = 420$. Then,  we have to sum over all triples, taking into account that we have to divide by two because the order of the pairs does not matter. To be careful with the expressions we are going to define:
\begin{align}
   A_{ijk} &= 210 \sum_{i,j,k=1, j\neq i, k\neq i,k\neq j}^\infty T_{4,i,i,i,i}T_{2,j,j}T_{2,k,k}Q_i^4(t)Q_j^2(t)Q_k^2(t) \\
   \nonumber
   A_{ijk} &= A_{ijk} + \delta_{kj}(A_{ijk} - A_{ijk}) + \delta_{ki}(A_{ijk} - A_{ijk}) = A_{ijk}(1 +\delta_{kj} +\delta_{ki}) - (\delta_{kj}+\delta_{ki})A_{ijk} \\ \label{termdivision}
   &= \underbrace{A_{ijk}(1 +\delta_{kj} +\delta_{ki})(1 + \delta_{ij})}_{(A)} -\underbrace{ (\delta_{kj}+\delta_{ki})A_{ijk}(1 + \delta_{ij})}_{(B)}-\underbrace{\delta_{ij}\big[A_{ijk}(1 +\delta_{kj} +\delta_{ki}) - (\delta_{kj}+\delta_{ki})A_{ijk}\big]}_{(C)} \,.
\end{align}
The first term (A) is: 
\begin{align}
     A_{ijk}&(1 +\delta_{kj} +\delta_{ki})(1 + \delta_{ij}) =210 \bigg(\sum_{i=1}^\infty T_{4,i,i,i,i}Q_i^4(t)\bigg) \bigg(\sum_{j=1}^\infty T_{2,j,j}Q_j^2(t)\bigg)^2 \;.
\end{align}
The second term (B) is: 
\begin{align}
     (\delta_{kj}+\delta_{ki})A_{ijk}(1 + \delta_{ij}) =210 \bigg(\sum_{i=1}^\infty T_{4,i,i,i,i}Q_i^4(t)\bigg) \bigg(\sum_{j=1}^\infty T_{2,j,j}^2 Q_j^4(t)\bigg)+210 \bigg(\sum_{i=1}^\infty T_{4,i,i,i,i}T_{2,i,i}Q_i^6(t)\bigg) \bigg(\sum_{j=1}^\infty T_{2,j,j}Q_j^2(t)\bigg).
\end{align}
The third term (C) is: 
\begin{align}
    \delta_{ij}&[A_{ijk}(1 +\delta_{kj} +\delta_{ki}) - (\delta_{kj}+\delta_{ki})A_{ijk}] \\\nonumber
    &= 210 \bigg(\sum_{i=1}^\infty T_{4,i,i,i,i}T_{2,i,i}Q_i^6(t)\bigg) \bigg(\sum_{j=1}^\infty T_{2,j,j}Q_j^2(t)\bigg)-210 \bigg(\sum_{i=1}^\infty T_{4,i,i,i,i}T_{2,i,i}^2Q_i^8(t)\bigg)-210\bigg(\sum_{i=1}^\infty T_{4,i,i,i,i}T_{2,i,i}^2Q_i^8(t)\bigg).
\end{align}
Therefore we have:
\begin{align}
   210& \sum_{i,j,k=1, j\neq i, k\neq i,k\neq j}^\infty T_{4,i,i,i,i}T_{2,j,j}T_{2,k,k}Q_i^4(t)Q_j^2(t)Q_k^2(t) \\
   &= 210\bigg[\bigg(\sum_{i=1}^\infty T_{4,i,i,i,i}Q_i^4(t)\bigg) \bigg(\sum_{j=1}^\infty T_{2,j,j}Q_j^2(t)\bigg)^2- 2 \bigg(\sum_{i=1}^\infty T_{4,i,i,i,i}T_{2,i,i}Q_i^6(t)\bigg) \bigg(\sum_{j=1}^\infty T_{2,j,j}Q_j^2(t)\bigg) \nonumber\\
   &\phantom{=,}-\bigg(\sum_{i=1}^\infty T_{4,i,i,i,i}Q_i^4(t)\bigg) \bigg(\sum_{j=1}^\infty T_{2,j,j}^2 Q_j^4(t)\bigg)+ 2\bigg(\sum_{i=1}^\infty T_{4,i,i,i,i}T_{2,i,i}^2Q_i^8(t)\bigg) \bigg] \;. \nonumber
\end{align}
The fifth term's coefficient comes from the possible combinations with 4 pairs of indices with $i,j,k,l$:
\mbox{${8 \choose 2}{6 \choose 2}{4 \choose 2}{2 \choose 2}=2520$}. Taking into account the symmetries---i.e. the order of the pairs does not matter---we have to divide by the total number of 4-permutations, which is 24. Then we have, naming this fifth term by $B_{ijkl}$ that
\begin{align}
    B_{ijkl} &= 105\sum_{i,j,k,l=1, j\neq i, k\neq i,j ,l\neq i,j,k}^\infty T_{8,i,i,j,j,k,k,l,l}Q_i^2(t)Q_j^2(t)Q_k^2(t)Q_l^2(t) \\
     B_{ijkl} &=  B_{ijkl}(1 + \delta_{li} +  \delta_{lj} + \delta_{lk})-B_{ijkl}(\delta_{li} +  \delta_{lj} + \delta_{lk})
\end{align}
Taking into account the result in Eq.~\eqref{termdivision} and the division in parts (A), (B) and (C) we obtain, for (A):
\begin{align}
   \big[B_{ijkl}(1 + \delta_{li} &+  \delta_{lj} + \delta_{lk})-B_{ijkl}(\delta_{li} +  \delta_{lj} + \delta_{lk})\big](1+\delta_{kj}+\delta_{ki})(1+\delta_{ji}) \\
   &= 105 \bigg[\bigg(\sum_{i=1}^\infty T_{2,i,i}Q_i^2(t)\bigg)^4 - 3 \bigg(\sum_{i=1}^\infty T_{2,i,i}^2Q_i^4(t)\bigg) \bigg(\sum_{j=1}^\infty T_{2,j,j}Q_j^2(t)\bigg)^2\bigg] \;. \nonumber
\end{align}
For (B) we have then
\begin{align}
    \big[B_{ijkl}&(1 + \delta_{li} +  \delta_{lj} + \delta_{lk})-B_{ijkl}(\delta_{li} +  \delta_{lj} + \delta_{lk})\big](\delta_{kj}+\delta_{ki})(1 + \delta_{ij})\\
    &=105\bigg[2\bigg(\sum_{i=1}^\infty T_{2,i,i}^2Q_i^4(t)\bigg) \bigg(\sum_{j=1}^\infty T_{2,j,j}Q_j^2(t)\bigg)^2-2\bigg(\sum_{i=1}^\infty T_{2,i,i}^2Q_i^4(t)\bigg)^2-4 \bigg(\sum_{i=1}^\infty T_{2,i,i}^3Q_i^6(t)\bigg) \bigg(\sum_{j=1}^\infty T_{2,j,j}Q_j^2(t)\bigg)\bigg] \;. \nonumber
\end{align}
For (C) we have
\begin{align}
    \big[&B_{ijkl}(1 + \delta_{li} +  \delta_{lj} + \delta_{lk})-B_{ijkl}(\delta_{li} +  \delta_{lj} + \delta_{lk})\big]\big[(1 +\delta_{kj} +\delta_{ki}) - (\delta_{kj}+\delta_{ki})\big]\delta_{ij}\\
    &=105\bigg[\bigg(\sum_{i=1}^\infty T_{2,i,i}^2Q_i^4(t)\bigg) \bigg(\sum_{j=1}^\infty T_{2,j,j}Q_j^2(t)\bigg)^2-4 \bigg(\sum_{i=1}^\infty T_{2,i,i}^3Q_i^6(t)\bigg) \bigg(\sum_{j=1}^\infty T_{2,j,j}Q_j^2(t)\bigg) \nonumber\\
    &\phantom{=105[[}-\bigg(\sum_{i=1}^\infty T_{2,i,i}^2Q_i^4(t)\bigg)^2 + 6\bigg(\sum_{i=1}^\infty T_{2,i,i}^4Q_i^8(t)\bigg)\bigg] \;. \nonumber
\end{align}
Finally,
\begin{align}
    B_{ijkl}=105\bigg[&\bigg(\sum_{i=1}^\infty T_{2,i,i}Q_i^2(t)\bigg)^4 - 6 \bigg(\sum_{i=1}^\infty T_{2,i,i}^2Q_i^4(t)\bigg) \bigg(\sum_{j=1}^\infty T_{2,j,j}Q_j^2(t)\bigg)^2 + 3\bigg(\sum_{i=1}^\infty T_{2,i,i}^2Q_i^4(t)\bigg)^2\\
    &\phantom{==}+8\bigg(\sum_{i=1}^\infty T_{2,i,i}^3Q_i^6(t)\bigg) \bigg(\sum_{j=1}^\infty T_{2,j,j}Q_j^2(t)\bigg)
    - 6\bigg(\sum_{i=1}^\infty T_{2,i,i}^4Q_i^8(t)\bigg)\bigg] \;. \nonumber
\end{align}
Joining everything,
\begin{align}
    \langle \hat{q}_\textsc{d}(t)^8\rangle &=\sum_{i=1}^\infty (T_{8,i,i,i,i,i,i,i,i}-28T_{6,i,i,i,i,i,i}T_{2,i,i} -35T_{4,i,i,i,i}^2+420T_{4,i,i,i,i}T_{2,i,i}^2-630T_{2,i,i}^4 )Q_i^8(t)\\
    &+ \bigg(\sum_{i=1}^\infty (28T_{6,i,i,i,i,i,i}-420 T_{4,i,i,i,i}T_{2,i,i}+840 T_{2,i,i}^3) Q_i^6(t)\bigg)\bigg(\sum_{j=1}^\infty T_{2,j,j}Q_j^2(t)\bigg) \nonumber  \\
    &+ \bigg(\sum_{i=1}^\infty T_{4,i,i,i,i} Q_i^4(t)\bigg)\bigg(\sum_{i=1}^\infty (35 T_{4,i,i,i,i}-210 T_{2,i,i}^2 )Q_i^4(t)\bigg) + \nonumber\\ 
    &+\bigg(\sum_{i=1}^\infty (210 T_{4,i,i,i,i}-630 T_{2,i,i}^2)Q_i^4(t)\bigg) \bigg(\sum_{j=1}^\infty T_{2,j,j}Q_j^2(t)\bigg)^2 +105\bigg[\bigg(\sum_{i=1}^\infty T_{2,i,i}Q_i^2(t)\bigg)^4 + 3\bigg(\sum_{i=1}^\infty T_{2,i,i}^2Q_i^4(t)\bigg)^2\bigg] \;. \nonumber
\end{align}
Finally, we note that for $i\neq 3, 4$ we have
\begin{align}
    &T_{8,i,i,i,i,i,i,i,i}-28T_{6,i,i,i,i,i,i}T_{2,i,i}, -35T_{4,i,i,i,i}^2+420T_{4,i,i,i,i}T_{2,i,i}^2-630T_{2,i,i}^4 =0,\\
    &28T_{6,i,i,i,i,i,i}-420 T_{4,i,i,i,i}T_{2,i,i}+840 T_{2,i,i}^3 = 0,\\
    &210 T_{4,i,i,i,i}-630 T_{2,i,i}^2 = 0.
\end{align}

\twocolumngrid

\bibliography{references}
\end{document}